\documentclass[aps,amsmath,twocolumn,amssymb,floatfix,showpacs,superscriptaddress,nofootinbib,longbibliography]{revtex4-1}
\usepackage{MnSymbol}
\usepackage{braket}
\usepackage[dvipsnames]{xcolor}
\usepackage{float}
\usepackage{subfigure}
\usepackage{tikz}
\usepackage{comment}
\usepackage{textcomp}
\usepackage{physics,array}
\usepackage[export]{adjustbox}

\usepackage[colorlinks=true,linktoc=page,linkcolor=blue,citecolor=blue,urlcolor=blue]{hyperref}

\newcommand{\vect}[1]{\boldsymbol{\mathrm{#1}}}

\mathchardef\mhyphen="2D

\newcommand\bea{\begin{eqnarray}}
\newcommand\eea{\end{eqnarray}}
\newcommand\beq{\begin{equation}} 
\newcommand\eeq{\end{equation}}

\usepackage[normalem]{ulem}
\definecolor{lime}{HTML}{A6CE39}
\usepackage{sidecap,tikz}
\DeclareRobustCommand{\orcidicon}{\hspace{-1.0mm}
	\begin{tikzpicture}
	\draw[lime, fill=lime] (0.0,0.0) 
	circle [radius=0.15] 
	node[white] {{\fontfamily{qag}\selectfont \tiny \,ID}};
	\draw[white, fill=white] (-0.0525,0.095) 
	circle [radius=0.007];
	\end{tikzpicture}
	\hspace{-3.0mm}
}
\definecolor{orchid4}{HTML}{804080}
\foreach \x in {A, ..., Z}{\expandafter\xdef\csname orcid\x\endcsname{\noexpand\href{https://orcid.org/\csname orcidauthor\x\endcsname}{\noexpand\orcidicon}}
}

\AtBeginDocument{
	\newwrite\bibnotes
	\def\bibnotesext{Notes.bib}
	\immediate\openout\bibnotes=\jobname\bibnotesext
	\immediate\write\bibnotes{@CONTROL{REVTEX41Control}}
	\immediate\write\bibnotes{@CONTROL{
			apsrev41Control,author="08",editor="1",pages="1",title="1",year="1"}}
	\if@filesw
	\immediate\write\@auxout{\string\citation{apsrev41Control}}
	\fi
}

\begin{document}

\renewcommand{\thefootnote}{\fnsymbol{footnote}}


\title{Deciphering competing interactions of Kitaev-Heisenberg-\texorpdfstring{$\Gamma$}~~system in clusters:\\part II - dynamics of Majorana fermions}	

\author{Sheikh Moonsun Pervez\orcidA{}}
\email{moonsun@iopb.res.in}
\affiliation{Institute of Physics, Sachivalaya Marg, Bhubaneswar-751005, India}
\affiliation{Homi Bhabha National Institute, Training School Complex, Anushakti Nagar, Mumbai 400094, India}

\author{Saptarshi Mandal\orcidB{}}
\email{saptarshi@iopb.res.in}
\affiliation{Institute of Physics, Sachivalaya Marg, Bhubaneswar-751005, India}
\affiliation{Homi Bhabha National Institute, Training School Complex, Anushakti Nagar, Mumbai 400094, India}
\begin{abstract}
We perform a systematic and exact study of Majorana fermion dynamics in the Kitaev-Heisenberg-$\Gamma$ model in a few finite-size clusters increasing in size up to twelve sites. We employ exact Jordan-Wigner transformations to evaluate certain measures of Majorana fermion correlation functions, which effectively capture matter and gauge Majorana fermion dynamics in different parameter regimes. An external magnetic field is shown to produce a profound effect on gauge fermion dynamics. Depending on certain non-zero choices of other non-Kitaev interactions, it can stabilise it to its non-interacting Kitaev limit. For all the parameter regimes, gauge fermions are seen to have slower dynamics, which could help build approximate decoupling schemes for appropriate mean-field theory. The probability of Majorana fermions returning to their original starting site shows that the Kitaev model in small clusters can be used as a test bed for the quantum speed limit.
\end{abstract}
\date{\today}
\maketitle
\section{Introduction}\label{introduction}
For the last couple of decades, Kitaev model~\cite{kitaev-2006} has established itself as a paradigmatic model which motivated various aspects of spin-liquid physics~\cite{savary-review,Balents2010,hermanns-2018,Loidl2021,simon-2022} and many body aspect of strongly correlated condensed matter system in general. These studies can be grouped into different categories. One is the non-interacting aspect of Kitaev physics including exact fractionalization of spins ~\cite{smandal-2007,subhajit-defect}, the effect of disorder and localization~\cite{kao-2021,nasu-2020}, entanglement studies~\cite{Qi-2010,mandal-2016,naveen-2018}  and extension to different lattices ~\cite{kivelson-2007,vala-csl-2010,naveen-2008,eschmann-2020}, topological degeneracies~\cite{mandal-toriccode,mandaljpa}, driven Kitaev model~\cite{subhajit-defect,Sasidharan_2024}, to mention a few. The second category includes a realistic effort to consider more generalized spin-Hamiltonian which includes non-Kitaev interactions such as external magnetic field~\cite{tikhonov-2011,lunkin-2019}, Ising term~\cite{mandal-subhro-2011}, Heisenberg term~\cite{chalaupka-2010,animesh-2020,animesh-2021, knolle-2018} and more general interactions popularly known as $J\mhyphen K\mhyphen\Gamma$ model~\cite{Rau-2014}. Each of these studies has a common aim: to chart the exact parametric dependency where the Kitaev-spin liquid phase can be unambiguously characterized. The emergence of orbital magnetization in Kitaev magnets~\cite{saikat-lin}, and the possibility of tuning the exchange parameters by using light to stabilize the spin liquid phase~\cite{Kumar2022}, observe inverse Faraday effect~\cite{saikat-inverseFARADAY} have been explored.
\\\indent
Apart from the above theoretically motivated works, contemporarily,  a handful of materials have also been proposed to possess Kitaev-like interaction along with other kinds of interactions ~\cite{chalaupka-2010,simon-2022,takagi-2019,motome-2020,hermanns-2018,abanerjee-2016}. The salient feature of these materials is that non-Kitaev interactions make the coveted spin-liquid state absent at very low temperatures. However, at a relatively intermediate higher temperature, the Kitaev physics - fractionalization of spins into Majorana fermions and fluxes, along with their intrinsic differences in the length and time scale of their dynamics - enable certain aspects of Kitaev spin liquid state to be realized experimentally~\cite{janssen-2017,chern-2021,czajka-2021,niravkumar-2019,wulferding-2020,berke-prb-2020,haoxiang-nature-2012,balz-prb-2021,plessis-2020}. Further recent studies show that proximity effect of $\alpha\mhyphen{\rm RuCl}_3$ in graphene heterostructure~\cite{sananda-2019,sananda-2021} increases the strength of Kitaev interaction and induces an insulator to metal transition which might lead to exotic superconducting states. Such heterostructure study holds great promise for interesting application-oriented Kitaev-motivated physics to be observed in the near future.
\\
\indent
Though a lot of work has been done to explore various aspects of the Kitaev Model, as mentioned earlier, only a limited theoretical works~\cite{knolle-2018,nasu-motome-2019,nandini-2019}  attempted to investigate the dynamics of Majorana fermions (matter and gauge) for a generalized Kitaev model. In a previous study~\cite{knolle-2018}, matter and gauge fermion's time-dependent correlations have been investigated for $J\mhyphen K\mhyphen \Gamma$ model using an augmented patron theory, which is limited only to few choices parameters, and effect of magnetic field has not been considered. On the other hand, though a magnetic field has been considered in~\cite{nandini-2019}, the study only considers a Kitaev model in the absence of Heisenberg and $\Gamma$ interactions. Similarly, in the study of non-equilibrium dynamics of Majorana fermions~\cite{nasu-motome-2019}, only Kitaev interaction is considered. This motivates us to explicitly investigate the dynamics of matter Majorana fermion and gauge Majorana fermions in detail within a single study for all possible parameter regions and find the effect of a magnetic field.
\\
\begin{figure}
	\centering
	\includegraphics[width=1\columnwidth,height=!]{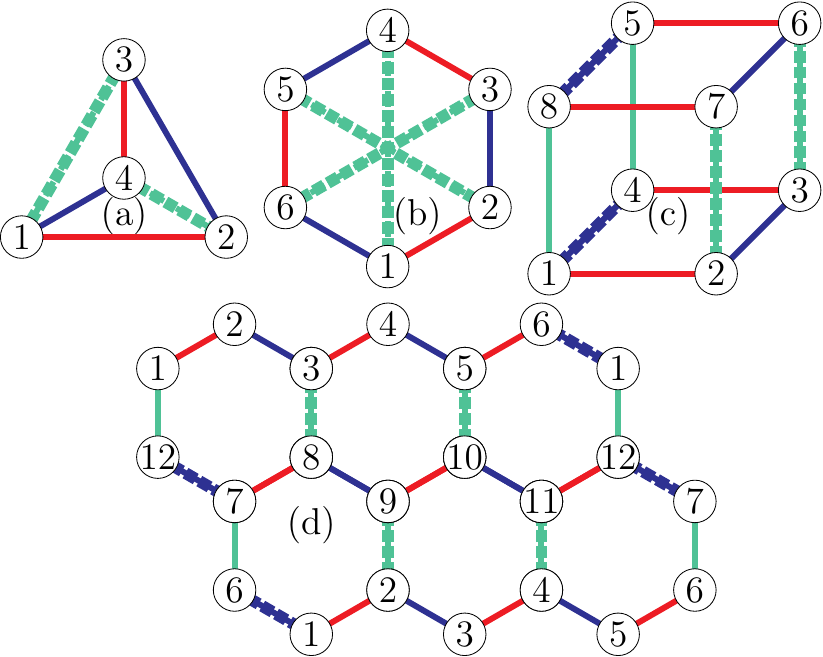}
	\caption{Depiction of (a) 4-site, (b) 6-site, (c) 8-site, and (d) 12-site Kitaev cluter. A bond's red, blue, and green colour indicates $x,~y,$ and $z$-type Ising interaction between the two sites that hold that particular bond. Thickly dotted bonds are `normal bonds', and the other bonds are `tangent bonds', as described in the main text.}
	\label{clusterfig_2}
\end{figure}
\indent
To this end, we have considered various small clusters, starting from four-site to twelve-site, for this purpose. The schematics of the clusters are shown in FIG.\ref{clusterfig_2}. Though we anticipate that the Majorana dynamics may contain significant boundary effects in small clusters, studying the interplay of Majorana dynamics in small clusters is an interesting theoretical question that deserves attention. Especially the question we ask: What is the effect of competing interactions in a generalized $J\mhyphen K\mhyphen \Gamma$ model, and how does the competition manifest in Majorana dynamics, both in matter and gauge sectors and what is the role of external magnetic field on these. To our surprise, in our exact numerics, we have explored interesting aspects such as magnetic field can potentially make the gauge fermions conserved and behave similarly to the non-interacting limit. Further, we found that certain aspects of quantum speed limit can be studied in these systems. The presentation in this paper is organized as follows. First, we introduce the various clusters we considered, the exact Jordan-Wigner transformation used, and the order parameter considered in section~\ref{kitaev-clusters-and-JW-transformation}. The result of gauge and matter fermions dynamics are described in section~\ref{section-gauge-majorana-dynamics}. We conclude with a discussion in section~\ref{discussion}.
\section{Kitaev-cluster-Basics and Jordan-Wigner fermionization}\label{kitaev-clusters-and-JW-transformation}
\begin{figure}\centering
	\includegraphics[width=1\columnwidth,height=!]{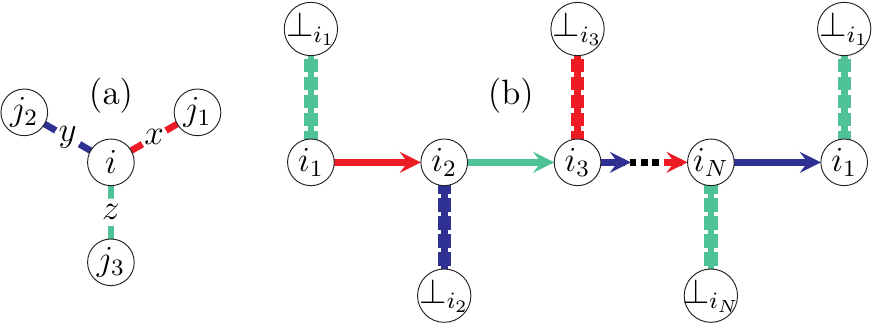}
	\caption{(a) Basic constituent of Kitaev-type cluster; a site is connected with three neighbours with direction-dependent Ising-like interaction. (b) Typical Jordan-Wigner path for fermionization. Solid lines are tangent bonds, and dotted lines depict normal bonds.}
	\label{JW-path_2}
\end{figure}
The defining Hamiltonian we investigate in our study is the $J\mhyphen K\mhyphen \Gamma$ model \cite{Rau-2014,wang-2021}. We work in isotropic Kitaev limit ($K_x=K_y=K_z= K $) and fix the energy scale by setting $|K|=1$. We use natural units, i.e., Planck constant $\hbar=1$, and Boltzmann constant $k_B=1$. In these units (and absorbing a 1/2-factor (per spin) in the parameters), the Hamiltonian becomes,
\begin{eqnarray}\label{hamiltonian}
H=H_K+H_J+H_\Gamma+H_Z.
\end{eqnarray}
Where $H_K$ is the Kitaev part of the Hamiltonian. It is direction-dependent Ising interaction on each bond and is given by,
\begin{eqnarray}\label{hamiltonian-kitaev-part}
H_K= K \sum_{{\langle j, k \rangle}_x} \sigma_{j}^{x} \sigma_{k}^{x}+ K \sum_{{\langle j, k \rangle}_y} \sigma_{j}^{y} \sigma_{k}^{y}+ K \sum_{{\langle j, k \rangle}_z} \sigma_{j}^{z} \sigma_{k}^{z},
\end{eqnarray}
where $\vect{\sigma}_j=\{\sigma_{j}^{x},\sigma_{j}^{y},\sigma_{j}^{z}\}$ are the spin 1/2 Pauli operators at $j^{\text{th}}$ site. The second term in Hamiltonian is the Heisenberg exchange interaction,
\begin{eqnarray}\label{hamiltonian-heisenberg-part}
H_J=J\sum_{\langle j, k \rangle} \vect{\sigma}_j\cdotp\vect{\sigma}_k,
\end{eqnarray}
and it is parameterized by the exchange coupling parameter $J$. The third term is an off-diagonal interaction, widely known as the $\Gamma$ term  given by,
\begin{eqnarray}\label{hamiltonian-gamma-part}
H_\Gamma=\Gamma\sum_{{\langle j, k \rangle}_{\alpha\neq\gamma\neq\delta}} \sigma_{j}^{\gamma} \sigma_{k}^{\delta}+\sigma_{k}^{\gamma} \sigma_{j}^{\delta}.
\end{eqnarray}
The last term in Hamiltonian denotes the effect of the external Zeeman field taken along $z$-direction. It is well known that a pure Kitaev model is described by a non-interacting Majorana fermion hopping problem ($H= \sum_{jk} J_{jk} i c_j c_k $), in the background of conserved $Z_2$ gauge fields $u_{ij}= ib_i b_j$ \cite{kitaev-2006}. In the appropriate representation, these conserved gauge field operators can be re-written as an equivalent number operator as $u_{ij}=2 \chi^{\dagger} \chi -I$, named bond-fermions \cite{smandal-2007,mandaljpa}. The Majorana operator ($c_i$) can also be appropriately recast in terms of complex fermion $\psi$, which on some bonds appear as $2 \psi^{\dagger} \psi-I$. The non-Kitaev interactions in the Hamiltonian given in Eq.\ref{hamiltonian} make the gauge fields acquire dynamics as they are no longer conserved. Quantitative analysis of these matter and gauge fields' dynamics are essential ingredients to understand the possible spin-fractionalization leading to spin-liquid phase appearing at intermediate temperature \cite{berke-prb-2020,nasu-motome-2015,koga-nasu-2019,li-2021-ncom}. This is also intimately connected to confinement-deconfinement transitions associated with the magnetically ordered state to quantum paramagnetic states, including the Kitaev spin liquid state.
\\\\\indent
The original solution of Kitaev \cite{kitaev-2006} suffers from enlargement of Hilbert space, and to avoid that, here we implement Jordan-Wigner transformation (JWT), as outlined in a previous study \cite{Mandal_2012}. Below, we briefly describe the salient features of the implementation of JWT.
\\\\\indent
{\it Fermionization procedure (`fermionic decoupling'):} The one-dimensional path for JWT is taken along $1\rightarrow2\rightarrow3\rightarrow\cdots\rightarrow N-1\rightarrow N\rightarrow1$ (FIG.~\ref{JW-path_2}), and bonds on this path are called tangent bonds; the bonds which are not on this path, are normal bonds (dotted bonds in FIG.~\ref{clusterfig_2} and FIG.~\ref{JW-path_2}). To do JWT, we define the Jordan-Wigner string operator $\mu_{i_m}$ as,
\begin{equation}
\mu_{i_m}=\prod_{j=1}^{m-1} \hat{n}_{i_j}\cdotp\vect{\sigma}_{i_j},
\end{equation}
where, $\hat{n}_{i_j}$ is the normal direction from the $j^{\text{th}}$ site, denoted as $i_j$. JW Fermions at site $i_m$ are ,
\begin{eqnarray}
\eta_{i_m}=(\hat{t}_{1,i_m}\cdotp\vect{\sigma}_{i_m})~\mu_{i_m},\nonumber\\
\xi_{i_m}=(\hat{t}_{2,i_m}\cdotp\vect{\sigma}_{i_m})~\mu_{i_m},
\end{eqnarray}
where, $\hat{t}_{1(2),i_m}$ is the in(out)-going tangent direction at site $i_m$ while walking along the JW path. Note that $\eta_{i_m}$ and $\xi_{i_m}$ are essentially Majorana fermions and note that, alternatively, one can express $\sigma$'s in terms of $\eta,\xi$ as shown below,
\begin{eqnarray}
\hat{t}_{1,i_m}\cdotp\vect{\sigma}_{i_m}=\eta_{i_m}\mu_{i_m}; \quad \hat{t}_{2,i_m}\cdotp\vect{\sigma}_{i_m}=\xi_{i_m}\mu_{i_m};\nonumber\\
i~\eta_{i_m}~\xi_{i_m}=i~(\hat{t}_{1,i_m}\cdotp\vect{\sigma}_{i_m})~(\hat{t}_{2,i_m}\cdotp\vect{\sigma}_{i_m})=\hat{n}_{i_m}\cdotp\vect{\sigma}_{i_m}.
\end{eqnarray}
With these definitions, spin-spin interaction in Hamiltonian is written in terms of $\eta,~\xi$, and the trailing operator $\mu$. The trail operator for the end bond ($N\rightarrow1$) is $S=\prod_{j=1}^{N} \sigma_{i_j}^{\text{n}_{\bot_{i_j}}}$. 
We combine the Majorana operators to define the complex fermions on normal bonds as $\eta_{i_k}=\psi_{i_k}+\psi_{i_k}^\dagger; ~~~ \eta_{\perp_{i_k}}=\frac{1}{i}(\psi_{i_k}-\psi_{i_k}^\dagger)$ and $\xi_{i_k}=\chi_{i_k}+\chi_{i_k}^\dagger;  ~~~ \xi_{\perp_{i_k}}=\frac{1}{i}(\chi_{i_k}-\chi_{i_k}^\dagger)$. With these definitions of $\eta$'s and $\xi$'s, the terms in Hamiltonian are written in terms of $\chi$ and $\psi$ and in general it is an interacting Hamiltonian. We follow exact diagonalization in suitable occupation number representations of $\chi$ and $\psi$ fermions and obtain the exact eigenstates to analyze the Majorana dynamics.
\\\\\indent
{\it Dynamics of Majorana fermions:} To investigate the fermion dynamics, we define  the  density-density correlation as,
\begin{eqnarray}\label{define_gt}
g(t)&=&\langle \text{GS}|e^{iHt} \hat{N}_i e^{-iHt}\hat{N}_i|\text{GS}\rangle,
\end{eqnarray}
where $\ket{\text{GS}}$ is the ground state of the system, and the number operator $\hat{N}_i$ is defined as,
\begin{eqnarray}
\hat{N}_i&=&2~\chi_{i}^\dagger\chi_{i}^{}-I~~~\text{for Gauge fermions},\nonumber\\
&=&2~\psi_{i}^\dagger\psi_{i}^{}-I~~~\text{for Matter fermions},
\end{eqnarray}
for $i=\left[1,\frac{N}{2}\right]$. We note that $g(t)$, in general, is a complex number, and we plot its absolute value. This can be simplified further considering $\hat{N}_i\ket{\rm GS}=\sum_{n}c_n\ket{n}$ where $\ket{n}$ denotes the $n^{\rm th}$ eigenstate, yielding $g(t)=\sum_{n} |c_n|^2 e^{-i(E_n-E_0)t}$. 
Here $E_n$ is the energy corresponding to $\ket{n}$ and  $E_0$ is the ground state energy, and $c_n=\bra{n}\hat{N}_i\ket{\rm GS}$. Physically, $g(t)$ measures how the {\it fermionic density} changes over time under unitary evolution at a given site, with respect to its initial value $g(0)$. The initial value has been effectively normalized to 1.
\\\\\indent
We have chosen $\hat{N}_i$ of the above form because they have a similar form of the conserved operator $u_{ij}$ associated with a given bond. However we notice that, $\langle \hat{N}_{i}^{(\psi)}(t) \hat{N}_{i}^{(\psi)}(0) \rangle= \langle (2\psi^{\dagger}_i \psi^{~}_i(t) -I)( 2\psi^{\dagger}_i \psi^{~}_i(0) -I) \rangle = \langle 4 n^{(\psi)}_{i}(t) n_{i}^{(\psi)}(0) - 2 n_{i}^{(\psi)}(t) - 2 n_{i}^{(\psi)}(0) + I \rangle $. Where we have used $n_{i}^{(\psi)}= \psi^{\dagger}_i \psi^{~}_i$. We note that $\langle n_{i}^{(\psi)} (t) \rangle= \langle n_{i}^{(\psi)} (0) \rangle$. Thus, effectively, 
$g(t) $ dependends on $ \langle n_{i}^{}(t) n_{i}^{} (0) \rangle $. We note that $n_{i}^{(\psi)}$ ($n_{i}^{(\chi)}$) acts as a projector on the ground state such that it only projects states with $\psi$ ($\chi$) fermion on the $i^{\rm th}$ bond. Thus, $g(t)$ yields the probability amplitude of those states to return to itself. In pure Kitaev limit, as $\hat{N}_{i}^{(\chi)}$ is static, $g_{\chi}(t)$ is constant in time but acquires time dependence once non-Kitaev terms are included. In the presence of these other interactions, $\psi$ and $\chi$ interact; thus, estimating those quantities in Eq.~\ref{define_gt} yields an understanding of the dynamics associated with non-Kitaev interactions. The next section summarises the salient characteristics of the gauge and matter Majorana fermion dynamics.
\section{Dynamics of Gauge and Matter Sector}\label{section-gauge-majorana-dynamics}
\begin{figure*}
	\includegraphics[width=2\columnwidth,height=10cm,keepaspectratio]{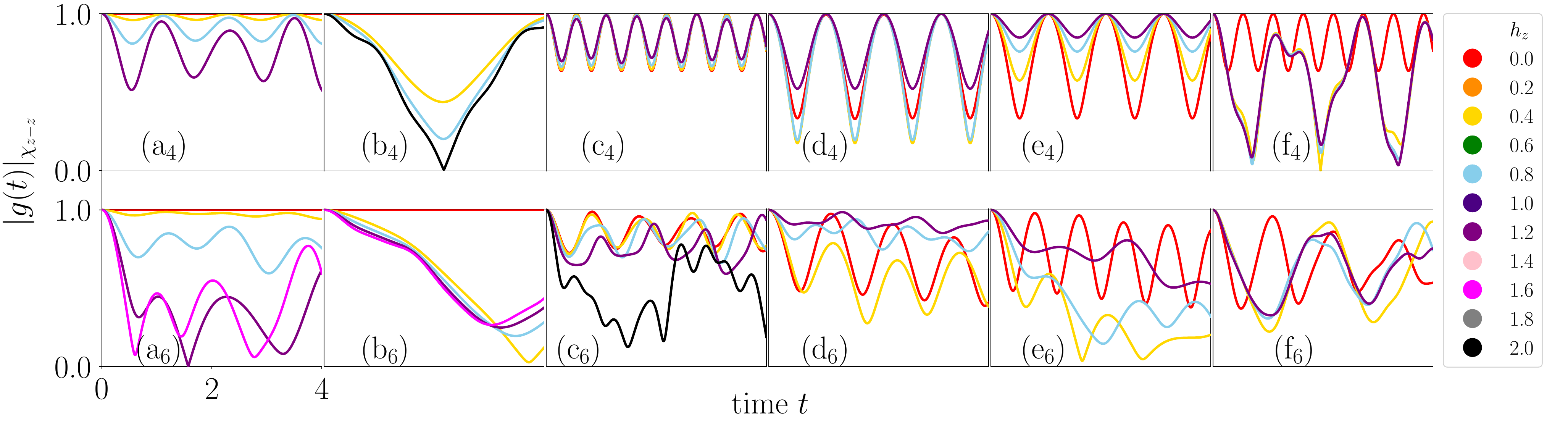}
	\caption{Gauge Majorana dynamics for $\Gamma=0$ and (a$_N$) $K=1.0,~J=0.0$, (b$_N$) $K=-1.0,~J=0.0$,  (c$_N$) $K=1.0,~J=0.5$, (d$_N$) $K=-1.0,~J=0.5$, (e$_N$) $K=1.0,~J=-0.5$, (f$_N$) $K=-1.0,~J=-0.5$. Different strengths of $h_z$ are given in the legend box.}
	\label{gauge-majo-dyn}
\end{figure*}
\begin{figure*}
	\includegraphics[width=2\columnwidth,height=10cm,keepaspectratio]{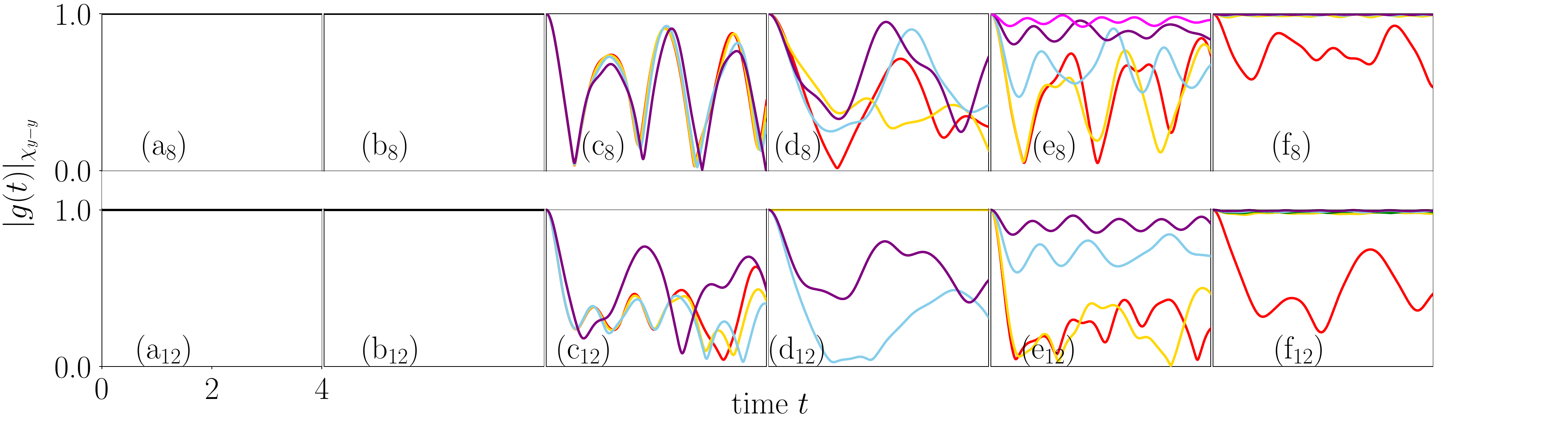}
	\caption{Gauge Majorana dynamics for $\Gamma=0$ and (a$_N$) $K=1.0,~J=0.0$, (b$_N$) $K=-1.0,~J=0.0$,  (c$_N$) $K=1.0,~J=0.5$, (d$_N$) $K=-1.0,~J=0.5$, (e$_N$) $K=1.0,~J=-0.5$, (f$_N$) $K=-1.0,~J=-0.5$. Different strengths of $h_z$ are given in the legend box of FIG.\ref{gauge-majo-dyn}.}
	\label{gauge-majo-dyn1}
\end{figure*}
\begin{figure*}
	\includegraphics[width=2\columnwidth,height=10cm,keepaspectratio]{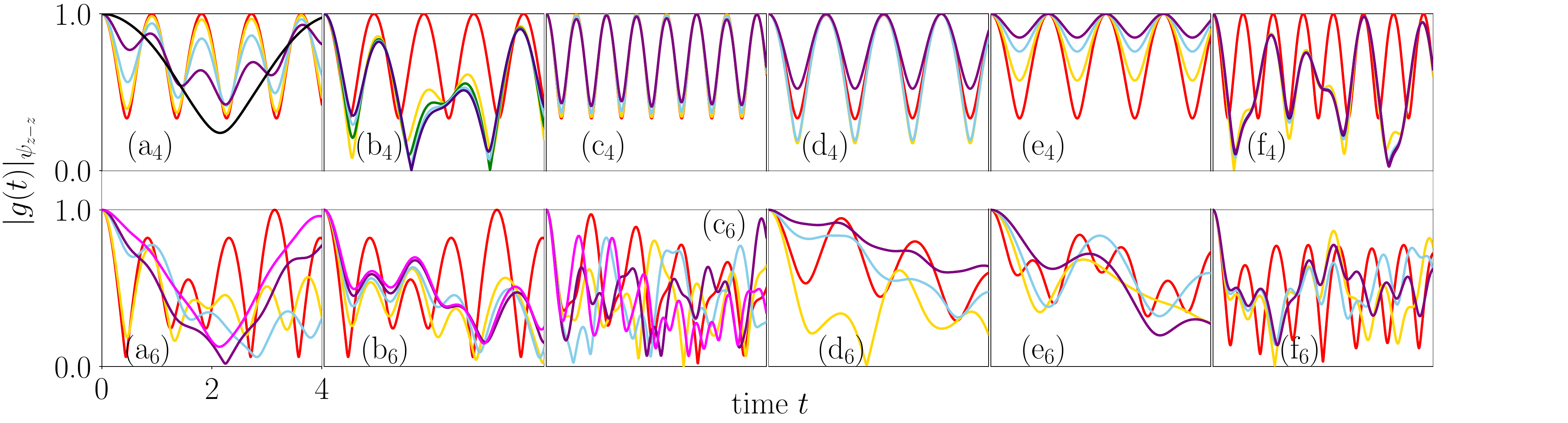}
	\caption{Matter Majorana dynamics for $\Gamma=0$ and (a$_N$) $K=1.0,~J=0.0$, (b$_N$) $K=-1.0,~J=0.0$,  (c$_N$) $K=1.0,~J=0.5$, (d$_N$) $K=-1.0,~J=0.5$, (e$_N$) $K=1.0,~J=-0.5$, (f$_N$) $K=-1.0,~J=-0.5$. Different strengths of $h_z$ are given in the legend box of FIG.\ref{gauge-majo-dyn}.}
	\label{matter-majo-dyn}
\end{figure*}
\begin{figure*}
	\includegraphics[width=2\columnwidth,height=10cm,keepaspectratio]{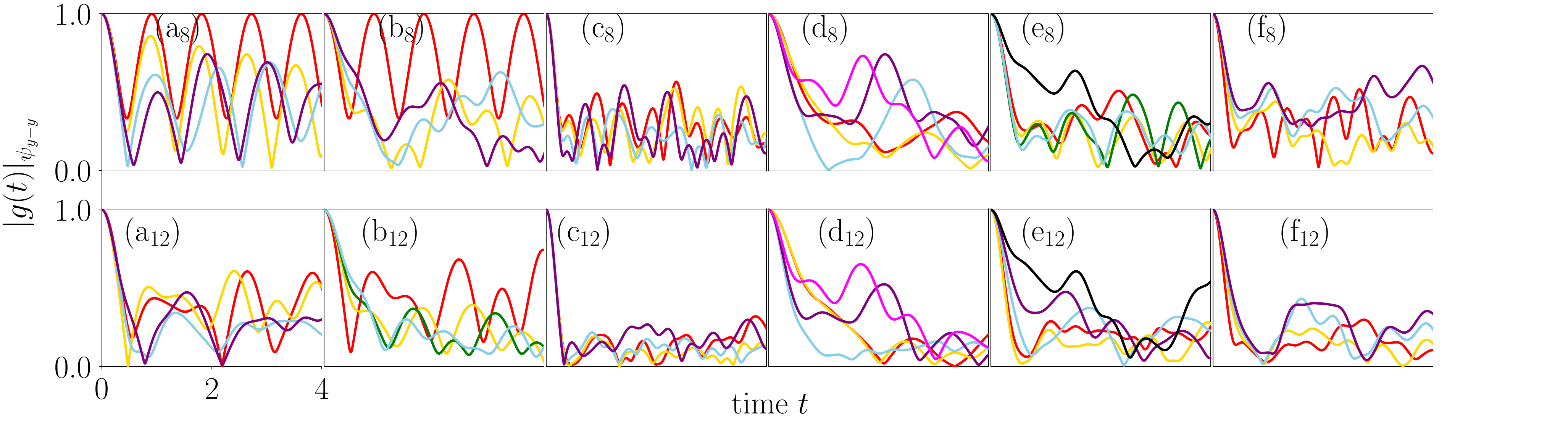}
	\caption{Matter Majorana dynamics for $\Gamma=0$ and (a$_N$) $K=1.0,~J=0.0$, (b$_N$) $K=-1.0,~J=0.0$,  (c$_N$) $K=1.0,~J=0.5$, (d$_N$) $K=-1.0,~J=0.5$, (e$_N$) $K=1.0,~J=-0.5$, (f$_N$) $K=-1.0,~J=-0.5$. Different strengths of $h_z$ are given in the legend box of FIG.\ref{gauge-majo-dyn}.}
	\label{matter-majo-dyn1}
\end{figure*}
\begin{figure}[h]
	\includegraphics[width=\columnwidth,height=!]{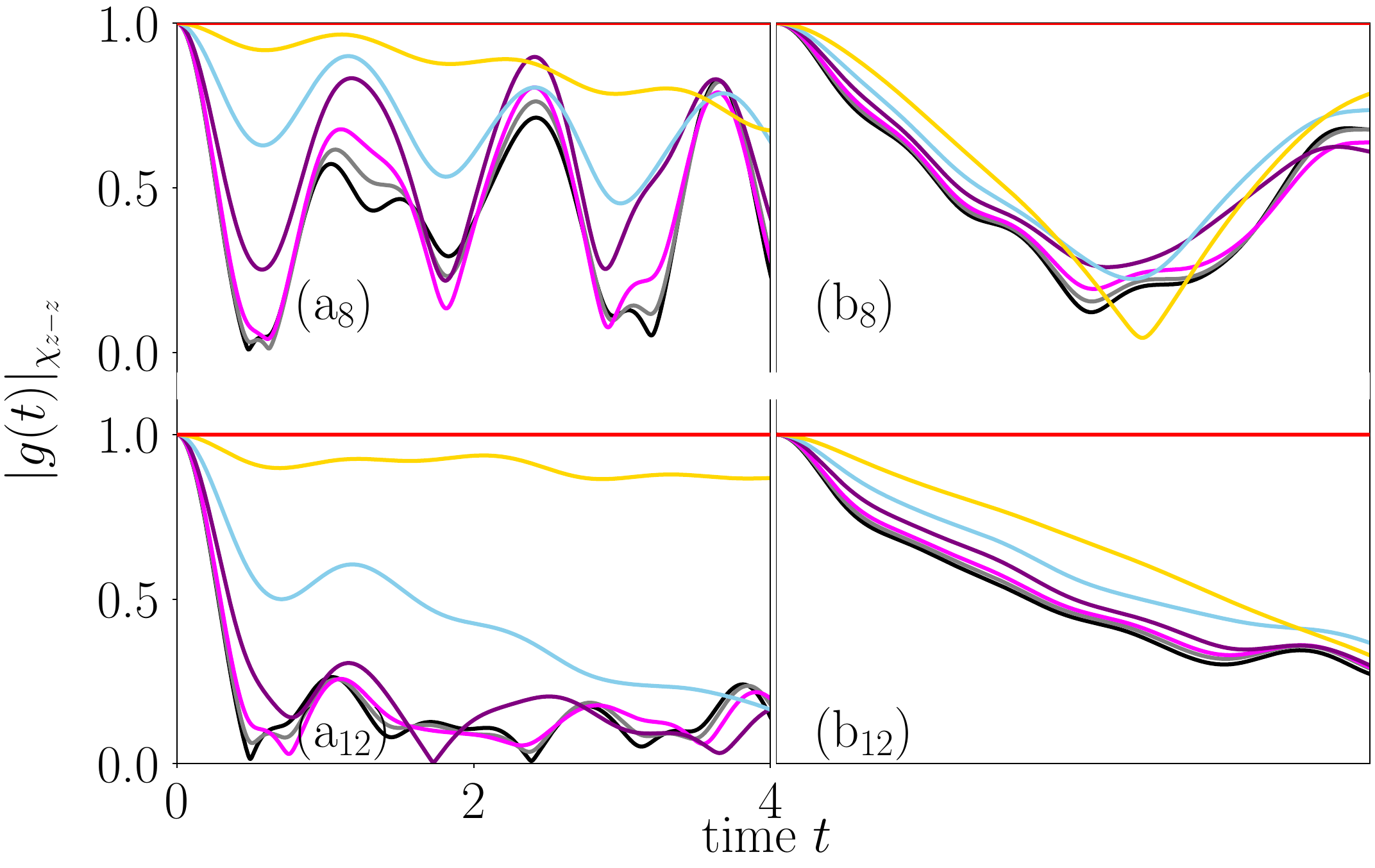}
	\caption{Evolution of gauge fermion defined on a $z$-type bond in (a$_8$, b$_8$) 8-site, and in (a$_{12}$,b$_{12}$) 12-site cluster. Parameter values are, $J=0$, $\Gamma=0$, along with (a$_{N}$) $K=1$ and (b$_{N}$) $K=-1$. Colour coding for different magnetic fields is the same as given in the legend box of FIG.\ref{gauge-majo-dyn}.}
	\label{gauge_Majorana_on_z_type_bond}
\end{figure}
\begin{figure}[H]
	\includegraphics[width=\columnwidth,height=!]{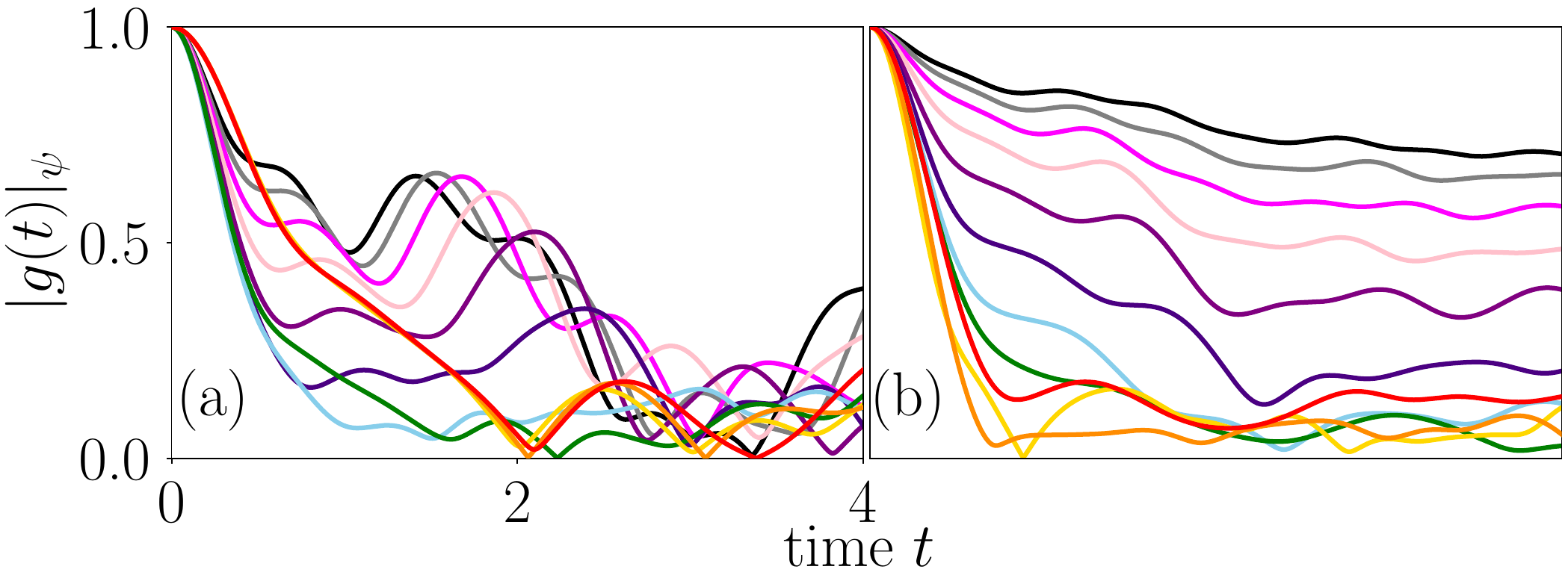}
	\caption{Evolution of matter sector which is defined on (a) bond (1-6: a $y$-type bond) and (b) bond (2-9: a $z$-type bond), for $K=-1$, $J=0.5$, and $\Gamma=0$, in the 12-site cluster. Different strengths of $h_z$ are the same as shown in the legend box of FIG.\ref{gauge-majo-dyn}.}
	\label{matter_majorana_comparison}
\end{figure}

To describe the dynamics of Majorana fermions, we consider bonds (1,3), (1,4), (1,4), (1,6) for 4, 6, 8, and 12-site cluster (`normal bonds' \cite{mandaljpa} connected with the first site). Thus, in the text below, whenever we mention $\chi$ or $\psi$ fermion, it is to be understood that they are defined on these bonds. Occasionally, for the 8-site and 12-site clusters, we mention $\chi_2$, which are defined on bonds (2,7) and (2,9), respectively; these are fermions on a $z$-type normal bond (unlike the $y$-type normal bonds that are connected with the first site in these two clusters). This helps us to compare the results from the four and six-site clusters, which hold $z$-type normal bonds connected with the first site. Moreover, we describe the amplitude of $|g(t)|$ with respect to its value in the pure-Kitaev limit. That is, `zero amplitude' means $|g(t)|=1$, and `large amplitude' (away from $|g(t)|=1$, and closer to $|g(t)|=0$) means it has deviated from the Kitaev limit.
\\\\\indent
{\small {\it {Qualitative differences between four-site cluster and other clusters}}:}
In all of the parameter regimes, there are qualitative differences in the dynamics of gauge fermions between the four-site cluster and other larger clusters. For $\Gamma=0=J$, the dynamics of gauge fermions are very regular and periodic as we vary the external magnetic field. For intermediate AFM $J$, the oscillations are very regular. Amplitude decreases monotonously as we increase the strength of $h_z$ (see FIG.\ref{gauge-majo-dyn}, panel (e$_4$)). However, for FM $J$, appearances of additional frequency could be noticed (FIG.\ref{gauge-majo-dyn}, panel (f$_4$), which points out that the ground state is modified differently than the AFM case. In the 4-site cluster, when we do the ground state analysis, for $K=1,~J=-0.5$, gauge dynamics can be calculated analytically to get $|g(t)|=\frac{1}{3}(5+4\text{cos}(12Jt))^{1/2} $ at $h_z=0$. When $h_z$ is introduced, only diagonal elements of the Hamiltonian are modified, leading to an equally tractable analytic calculation to see that $\left|g(t)\right|=\left|\left|\alpha\right|^2+\left|\tilde{\alpha}\right|^2+4\left|\beta\right|^2e^{-12iJt}\right|$, where the ground state is represented in the form $\alpha\ket{\uparrow_1\uparrow_2\uparrow_3\uparrow_4}+\tilde{\alpha}\ket{\downarrow_1\downarrow_2\downarrow_3\downarrow_4}+\beta( \ket{\uparrow_1\downarrow_3}-\ket{\downarrow_1\uparrow_3}) (  \ket{\uparrow_2\downarrow_4}-\ket{\downarrow_2\uparrow_4}) $. The same expression for $g(t)$ is true for any $J=-K/2$ in the 4-site cluster in the absence of $\Gamma$.
\\\\\indent
The same qualitative difference discussed above can also be seen for matter fermions. However, we notice that, in general, the average amplitude of oscillations for gauge fermions is always larger than the matter fermions. This can be attributed to the non-local nature of the gauge fermions in terms of the spin operators and the fact that each flux sector is separated by an energy gap much larger than the fermionic spectrum for each flux sector.
\\\\\indent
{\small {\it {Regular oscillations and irregular oscillations}}:}
It is observed that depending on the $h_z$, gauge fermion oscillation could be regular to irregular type. For example, consider the 6-site cluster at $J=0$. Up to $h_z=0.6$, the oscillations are minimal, but after that, it suddenly increases\cite{yamada-2020} and reaches a maximum for $h_z=1.2$ and decreases again. This has been shown in FIG.\ref{gauge-majo-dyn}, panel (a$_6$). On numerous occasions, it has been observed that $h_z$ can reverse the response of gauge fermions amplitude or time period of oscillations. For the 8-site cluster, a different scenario is observed, comparing
$J=-0.5,~0.0,~0.5$. In FIG.\ref{gauge-majo-dyn1}, panel (e$_8$, f$_8$), for $J=-0.5$, the amplitude of oscillation decreases with the increase of $h_z$. For $J=0.5$ (FIG.\ref{gauge-majo-dyn1}, panel (c$_8$, d$_8$)), the oscillations remain more or less the same for all $h_z$; no special feature is seen. However, for the 12-site cluster, such a non-monotonous response is absent. Very interestingly, we see from FIG.\ref{gauge-majo-dyn1} panel (a$_8$, b$_8$, a$_{12}$, b$_{12}$) that the gauge fermion does not show any dependency on time for any $h_z$. This can be referred to as stabilizing gauge fields under an external magnetic field on certain bonds. This happens for gauge fermions defined on a $y$-type bond and when the magnetic field is applied along $\hat{z}$-direction. However, on a different bond, the gauge fermion dynamics is oscillatory, as shown in FIG.\ref{gauge_Majorana_on_z_type_bond}.
There, we have shown dynamics of $\chi_2$ gauge fermion, which is defined on a $z$-type bond, and it does show dynamics under the influence of magnetic field applied along the $\hat{z}$-direction. The above-controlled manipulation of gauge field dynamics could be of practical use.
\\\\\indent
The identical trends can also be seen for matter fermions, depending on the relative strength of $h_z$ and $J$. For example, one can compare $(N,K,J,h_z)$=$(6,1,0,0.4)$ and $(6,1,0,1.2)$. In the former case, many minimums are found, but for the latter, only a single minimum was found; see FIG.\ref{matter-majo-dyn}, panel (a$_6$). The nature of oscillation completely changed. Similar trends are there for $(N,K,J,h_z)$=$(6,-1,0.5,0)$, $(6,-1,0.5,0.4)$, $(6,-1,0.5,0.8)$, i.e., panel (d$_6$) in FIG.\ref{matter-majo-dyn} (for $\psi_2$ fermion (matter fermion defined on a different bond) too, we see equivalent trends). These three sets of parameters yield three different behaviours for the matter fermion. For 8-site cluster, such differences are also found for $(K, J)$=$(-1,-0.5)$, $(-1,0)$, $(-1,0.5)$ (see FIG.\ref{matter-majo-dyn1}, panel (b$_8$,d$_8$,f$_8$)), and also for $(K, J)$=$(1,0)$, $(1,0.5)$, $(1,-0.5)$ (panel (a$_8$,c$_8$,e$_8$)), where depending on external magnetic field, matter fermion's dynamics changes a lot. Similarly, for 12-site, the relevant set of parameters for such change in the nature of dynamics for matter fermion can be seen at $(K, J)$=$(1,-0.5)$, $(-1,0.5)$.
\\\\\indent
{\small {\it {Dependency of average amplitude and time period of oscillations on $h_z$ and $J$}}:}
Interestingly, it has been noticed that for $J=0$, the time period of oscillations is much larger for $K=-1$ than $K=1$ case; this can be noticed from the first and second columns of FIG.\ref{matter-majo-dyn} and FIG.\ref{matter-majo-dyn1}. In the first column, for finite $h_z$, the oscillations have more amplitude and time period than the second column. However, as we turn on finite $J$, the time period of oscillations becomes comparable, though it greatly depends on the relative strength of $K,~J$ and $h_z$. It is also observed that the external magnetic field helps revive the amplitude of matter fermion on that site. For example, consider the 12-site cluster where the rapid fluctuations are absent compared to smaller clusters. If we look at $(K,J)=(-1,0.5)$ (FIG.\ref{matter-majo-dyn1}, panel (d$_{12}$)), where in between $t=1$ and $t=3$, there is a revival of amplitude of matter fermions. Similarly we find for $(K, J)=(1,0)$ (panel (a$_{12}$) with $h_z=0.4$) $(K, J)=(\pm1,-0.5)$ (panel (e$_{12}$, f$_{12}$)). Similarly we see that for gauge fermions, $(K,J)=(-1,0)$ corresponds to larger time period in comparison to $(K,J)=(+1,0)$.
\\\\\indent
{\small {\it {Two different channels of dynamics of oscillations}}:}
It is also noted that depending on the initial position of the matter fermions, its time dependency could be very different. For comparison, we note the results for the 12-site cluster with the parameter value $(K, J,\Gamma)= (-1, 0.5,0)$, as shown in FIG.\ref{matter_majorana_comparison}.
We observe that whereas $\psi_1$ (defined on a $y$-type bond) decays rapidly, $\psi_2$ (defined on a $z$-type bond) has quite small oscillations and varies smoothly. The average magnitude of $g_{\psi}(t)$ corresponding to $\psi_2$ also gradually decreases consistently with increasing $h_z$. Depending on the initial position of the matter fermion, this different behaviour indicates local competition between different interactions, such as Heisenberg and Kitaev interaction. The same applies to FM $J$ with parameter $(K, J)=(-1,-0.5)$. There is a marked difference in the dynamics of $\psi_1$ and $\psi_2$. The same can be said for $J=0$. This observation could be beneficial for the controlled manipulation of Majorana fermions for useful quantum state operations.
\\\\\indent
{\small {\it {Stabilization of $Z_2$ gauge field by external magnetic field}} :} It has been observed that, as the applied magnetic field is in $z$-direction, it can stabilize the $Z_2$ (defined on a $z$-type bond) gauge field\cite{khwang-2022} as seen in for eight and twelve-site cluster ($K, J$)=($-1,-0.5$) (FIG.\ref{gauge-majo-dyn1}, panel (f$_8$,f$_{12}$). For intermediate value of positive $J$, the gauge field fluctuation decreases as we increase $h_z$ as seen for ($K,J$)= ($-1, 0.5$) (FIG.\ref{gauge-majo-dyn}, panel(d$_6$)). This can be thought of as stabilizing the spin-liquid or fractionalized state by an external magnetic field. For AFM $J$, the $Z_2$ gauge fields show more oscillatory patterns than FM $J$, where they become closer to unity.
\\\\\indent
{\it\small Gauge fermion dynamics in the plateau region of correlation}:
In our previous study \cite{pervez2023deciphering}, we have shown that in these clusters, in the presence of AFM $J$, the correlation function can show a plateau region where the ground state effectively does not change over a range of $J$ or $h_z$. So, it will be interesting to investigate the fate of gauge field dynamics in this plateau region.
\begin{figure}[H]
	\includegraphics[width=\columnwidth,height=!]{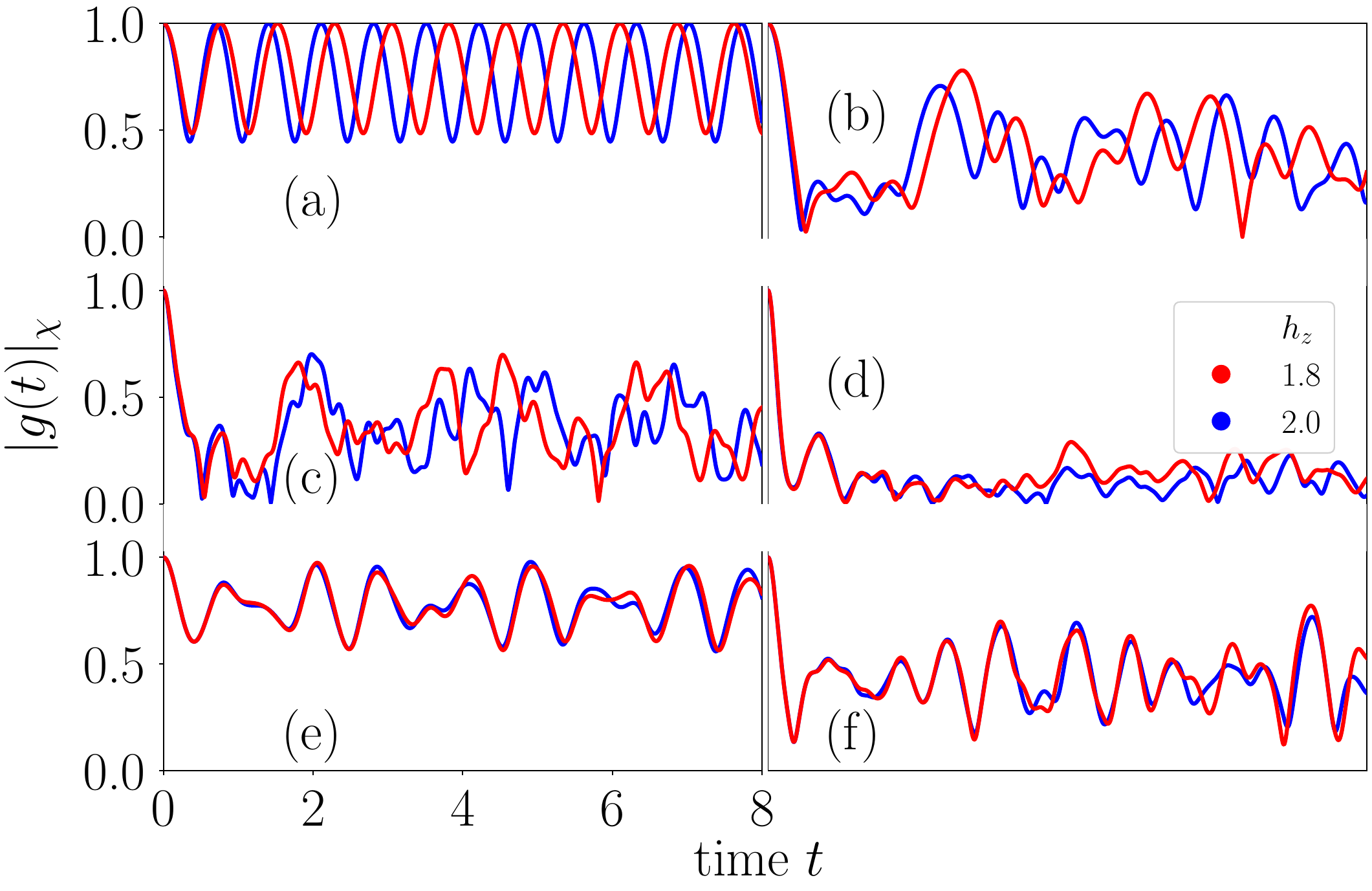}
	\caption{Evolution of gauge sector in the plateau region of correlation. (a) 4-site result; $\chi$ is defined on bond (1-3: a $z$-type bond), and $K=1$, $J=0.40$, $\Gamma=0$.  (b) 6-site result; $\chi$ is defined on bond (1-4: a $z$-type bond), and $K=1$, $J=0.13$, $\Gamma=0$. In 8-site, for $K=1$, $J=0.70$, $\Gamma=0$ (c) $\chi$ is defined on bond (2-7: a $z$-type bond), and (e)$\chi$ is defined on bond (1-4: a $y$-type bond). In 12-site, for $K=1$, $J=1.00$, $\Gamma=0$ (d) $\chi$ is defined on bond (2-9: a $z$-type bond), and (f)$\chi$ is defined on bond (1-6: a $y$-type bond). Different strength of $h_z$ is as shown in the legend box.}
	\label{gauge_majorana_plateau_region}
\end{figure}
In FIG.\ref{gauge_majorana_plateau_region}, we study $|g(t)|_\chi$ in the clusters for specific $J$ values as mentioned in the caption. At these $J$ values, for $h_z$= 1.8 and 2.0, correlation does not change for the respective cluster. As we observe, dynamics for $h_z=1.8$ is commensurate with that for $h_z=2.0$. Surprisingly, for $\chi$ fermion defined on a $y$-type bond (panel (e,f) of FIG.\ref{gauge_majorana_plateau_region}), the dynamics are almost identical for these two different strengths of $h_z$. This shows how the gauge fermions' dynamics really depend on the magnetic field's direction.
\\\\\indent
{\small {\it {Scenarios to define quantum speed limit (QSL)}} :} Recently, various aspects of the QSL, as defined by the evolution of a quantum state from a given state to a distinctly different state, attracted wide interest for various reasons \cite{deffner-2017,gal-ness-2022,pati-2022,arindam-2022}. In this context, we find that QSL could be investigated for the evolution of Majorana fermions. Initially, they begin with a given occupancy at a given bond. If the fermion occupation on that bond becomes zero at any future time, we can define QSL for our system. In this study of dynamics for gauge and matter fermions, we witness numerous occasions where such a situation occurs.
\\\\\indent
The fascinating occurrence of QSL happens for 4-site cluster with $(N, K, J,h_z)$=$(4,-1,0,2)$, $(4,-1,-0.5,0.4)$ (panel (b$_4$,f$_4$) of FIG.\ref{gauge-majo-dyn} and FIG.\ref{matter-majo-dyn}).
This happens in both the matter and gauge fermion sector. For the second set of parameters, they are remarkably repetitive with time. For AFM $K$, such realization does not happen (in the 4-site cluster) except for very large FM $J$. In 6-site cluster for the gauge fermions, we find $(N, K, J,h_z)$=$(6,1,0,1.2)$ being such an example (panel (a$_6$), FIG.\ref{gauge-majo-dyn}). For large AFM $J$, it also happens for matter fermions. There are occasions for gauge fermions on other bonds too for such phenomena, for example at $(N, K, J,h_z)= (8,-1,0,0.4)$ in $\chi_2$ gauge fermion (FIG.\ref{gauge_Majorana_on_z_type_bond}, panel (b$_8$)), and $(N,K,J,h_z)$=$(8,1,0,1.8)$, $(8,1,0,2.0)$ (FIG.\ref{gauge_Majorana_on_z_type_bond}, panel (a$_8$)). There are special cases where such phenomena are also observed for the 12-site cluster. A careful study and estimation of this QSL will be done in future studies as this is not the primary focus here.
\\\\\indent
{\small {\it {The main differences in the dynamics of Majorana fermions between twelve-site cluster and other clusters}} :} The most apparent difference between 12-site and other smaller clusters can be understood by analyzing $J=0$. In the absence of $J$, gauge fermion remains constant with time for clusters of all sizes when the external field is not there. However, matter fermions oscillate with fixed frequency and return to unity after characteristic time for four and eight-site clusters due to the backflow from the boundary. On the other hand, for the 6-site cluster, the matter fermions return to unity after a few local maxima. However, for the 12-site cluster, the matter fermions {\it never} return to unity (the time for returning is $t\sim5\times10^{2}$, which is two orders of magnitude larger than the other three smaller clusters). Secondly, for finite $J$, the oscillations in $\chi$ or $\psi$ fermions have a regular pattern for different $h_z$, but for smaller clusters, the pattern of dynamics varies greatly depending on $h_z$.
\\\\\indent
The appearance of multiple frequencies in the dynamics of Majorana fermions can be understood as follows. For a fixed gauge field configuration, the effective Hamiltonian for $\psi$ fermions reduces to a superconducting Hamiltonian with the presence of $ \eta_{ij} \psi_i \psi_j + h.c$. The ground state $\ket{\rm GS}$ is constructed as the vacuum of the quasi-particles where the states with an even number (including zero) of $\psi$ fermions appear with different probability amplitude. Now the operator $n_{i}^{(\psi)}= \psi^{\dagger}_i \psi^{~}_i$ selects (or projects) only those states with the dimer `$i$' being occupied.   These daughter states could be re-expressed as linear combinations of all the eigenstates (including excited and ground states) with different amplitudes. Thus the final expression of $g_{\psi}(t)$ contains the summation of $ \sum_{n} A_{n}\exp^{-i(E_n-E_0)t}$ where the detail analytic expression of $A_n$ is in general very complex. With a system size very large, substantial contribution to $A_{n}$ coming from numerous daughter states makes the probability amplitude oscillate with no fixed time period.
\\\\\indent
Now, for finite non-Kitaev interaction being present, the gauge fermions are no longer conserved, and hence, they also oscillate with time. Due to special symmetry, the matter and gauge fermion return to unity with identical frequency for the 4-site cluster. As the cluster size increases, the oscillations of matter and gauge fermions become irregular due to contributions from the increasing number of excited states for larger clusters. However, few universal characteristic features are present. One notes that, on average, the gauge fermion amplitude is always larger than the matter fermion. Also, matter fermions complete many cycles of oscillations within a given time window compared to gauge fermions. This points out that gauge fermions can still be approximated as slow varying background conserved operators for the matter fermions and hence may support the survival of the deconfined phase.
\subsection*{Dynamics of matter and gauge fermions at $\Gamma\neq 0$}
Next, we focus on the case where finite $\Gamma$ is present. For the sake of simplicity, we only describe the analysis in the absence of $J$. Corresponding plots are given in FIG.\ref{gauge_finite_Gamma},\ref{gauge_finite_Gamma1},\ref{matter_finite_Gamma},\ref{matter_finite_Gamma1}, in Appendix \ref{App:A_2}.
\\\\\indent
{\small {\it {Gauge fermion dynamics at $\Gamma=\pm0.5$}} :} When we compare the results of AFM $K$ (first and third column of FIG.\ref{gauge_finite_Gamma} and FIG.\ref{gauge_finite_Gamma1}) with that of FM one (second and fourth column), we observe that gauge fermions fluctuate more in FM $K$ case. However, crucial differences exist among the clusters. In 4 and 6-site, with increasing magnetic field, $|g(t)|$ gradually moves away from its initial value of $1$, whereas, in 8 and 12-site clusters, it moves closer to $1$. Also, when we notice that in the 4-site cluster, for FM $K$, $g(t)$ at high magnetic fields contain lower frequency and with lesser $h_z$, overall frequency is a little high (panel (b$_4$,d$_4$), FIG.\ref{gauge_finite_Gamma}). In the 6-site cluster, depending on the sign of $\Gamma$ and $K$, different $h_z$ mainly yield two different behaviours. For lower magnetic fields, $g(t)$ remains closer to its initial value, and at relatively higher $h_z$, $g(t)$ moves closer to zero. In larger clusters (8 and 12 sites), this kind of different behaviour, depending on $h_z$, is absent. We observe $g(t)$ vanishes immediately as we introduce $\Gamma$ (red line in FIG.\ref{gauge_finite_Gamma1}), which gets revived as we gradually increase $h_z$.
\\\\\indent
{\small {\it {Matter fermion dynamics at $\Gamma=\pm0.5$}} :} Here, we refer to FIG.\ref{matter_finite_Gamma} and FIG.\ref{matter_finite_Gamma1} of Appendix \ref{App:A_2}. Firstly, as we increase the system size, $g(t)$ gradually vanishes, even without a magnetic field. However, for a smaller cluster, there is a higher chance for the fermion to return to its initial state due to reflection. In addition, at lower $h_z$, fluctuation in $g(t)$ is more prominent for four and six-site than higher values of $h_z$. However, for larger clusters, an increase in $h_z$ only makes $g(t)$ away from zero. In the 12-site cluster (panel (a$_{12}$,b$_{12}$,c$_{12}$,d$_{12}$) of FIG.\ref{matter_finite_Gamma1}), the fermion never returns to its initial state. This happens because as the system evolves with time, the fermion gradually diffuses into the bulk, and the reflection from the boundary is ineffective in returning to its initial position because of the cluster's large size.
\section{Discussion}\label{discussion}
As mentioned earlier, the Kitaev model contains two types of Majorana fermions: matter Majorana fermions and gauge Majorana fermions. In the Kitaev limit, two gauge Majorana fermions combine to constitute a conserved $Z_2$ flux operator. On the other hand, matter Majorana fermions constitute a non-interacting hopping problem coupled to this conserved $Z_2$ field. However, with other non-Kitaev interactions, gauge fermions acquire dynamics. Various interesting consequences arise, such as the confinement-deconfinement transition, related to the revival of the spin-liquid phase at intermediate temperatures. The time evolution of matter and gauge fermions follow an oscillatory yet decaying pattern depending on different interactions. These oscillations have various interesting profiles, but their dependency on the external magnetic field is the most significant. The external magnetic field is seen to modulate significantly the amplitude of oscillations and also the average magnitude of it. Depending on the strength of the external magnetic field, it can freeze the gauge field without having any oscillations, showing remarkable dependence on the magnetic field.
\\\\\indent
In pure Kitaev limit, the gauge fields are static; effectively, they are {\it infinitely heavy}. With the application of magnetic field, they become {\it lighter}. A higher magnetic field makes the gauge fermion diffuse more into the bulk, as can be seen via a higher amplitude of $|g(t)|$. The frequency of oscillation in FM $K$ is smaller than AFM $K$. Next, we introduce non-zero Heisenberg strength and find that the dynamics are qualitatively similar for a competing $J$ and $K$. In both cases ($\pm K{\rm~and~}\mp J$), an increase in magnetic field results in a $|g(t)|$, which remains close to its initial value. When the sign of $J$ and $K$ are the same, the overall picture tells us that the increasing $h_z$ makes the amplitude of $|g(t)|$ reach closer to zero, indicating the fermionic density to be more diluted. Next, we looked at finite $\Gamma$ results in the absence of $J$. The gauge field dynamics are controlled mainly by $K$, with minimal effect from AFM or FM $\Gamma$. In the presence of $\Gamma$, increasing $h_z$ makes $|g(t)|$ to have higher amplitude for AFM $K$, whereas for FM $K$, a gradual increase in $h_z$ results in a $|g(t)|$ which approaches its initial value. In all cases, a bigger cluster shows slower dynamics, owing to the larger size, for it has more room for the fermion to get diluted. For some parameter values, the gauge field dynamics are seen to be stabilized by applying a magnetic field of proper magnitude and direction.
\\\\\indent
Next, we study the dynamics of matter fermions. A higher magnetic field on top of the Kitaev interaction case makes the matter fermion diffuse more. The frequency of oscillation in FM $K$ is smaller. Then we put non-zero Heisenberg strength and find out that, for a competing $J$ and $K$, just like the gauge dynamics, the matter dynamics are qualitatively similar here, too. In both cases, an increase in $h_z$ results in a $|g(t)|$ which approaches 1. Next, we study finite $\Gamma$ results in the absence of $J$. We find that the matter field dynamics are influenced mainly by $K$, with little respect towards AFM or FM $\Gamma$. In the presence of $\Gamma$, increasing $h_z$ makes the matter fermion dilute less into the bulk for FM $K$. In every case, a larger cluster shows slower dynamics.
\\\\\indent
On a different note, the 6-site cluster that we have considered here has a resemblance with the hexagonal-plaquette taken previously~\cite{alternative_6_site} (which contains next to next nearest neighbour interaction) after swapping a pair of sites (site-3 and site-5, to be specific) or by site-dependent gauge transformations. Thus, it is related to a physically motivated extended Kitaev model.
\\\\\indent
In the plateau region of correlation, where the ground state {\it almost} remains the same for a range of magnetic field (in $\hat{z}$-direction) values, for a given Heisenberg and Kitaev interaction, the gauge dynamics is {\it almost} identical when we look at the dynamics of a gauge field defined on a $y$-type bond. Meanwhile, gauge fermion, defined on a $z$-type bond, shows commensurate dynamics for multiple magnetic fields. An interesting realization of quantum speed limit(QSL) phenomena also appears where the probability of Majorana fermion vanishes from the initial value of unity. A detailed analysis of this QSL and its dependency on Heisenberg coupling and the external magnetic field will be followed in future studies.
\\\\\indent
With the recent intense effort to understand the Kitaev-Heisenberg-$\Gamma$ system, we think our extensive analysis would serve as a helpful reference. However, all the aspects found here may not be readily generalized to the thermodynamic system. Instead, it would be prudent to consider this an exciting platform to quantify the quantum and thermal fluctuations in the Kitaev-Heisenberg-$\Gamma$ system and use it to understand the thermodynamic system better. Further it may be interesting to examine Kitaev-Heisenberg model on other trivalent lattices\cite{naveen-2008,atanufisher,knollestar} to find quantitative differences in comparison to Honeycomb lattice which we leave for future study.
\section*{Acknowledgement} 	
S.M.P. acknowledges SAMKHYA (High-Performance Computing facility provided by the Institute of Physics, Bhubaneswar) for the numerical computation. S.M.P. also thanks Arnob Kumar Ghosh for useful discussions. S.M. acknowledges support from ICTP through the Associate's Programme (2020-2025). S.M. also thanks G. Baskaran and Nicola Seriani for the interesting discussions.
\bibliography{bibfile}{}

\begin{thebibliography}{62}%
\makeatletter
\providecommand \@ifxundefined [1]{%
 \@ifx{#1\undefined}
}%
\providecommand \@ifnum [1]{%
 \ifnum #1\expandafter \@firstoftwo
 \else \expandafter \@secondoftwo
 \fi
}%
\providecommand \@ifx [1]{%
 \ifx #1\expandafter \@firstoftwo
 \else \expandafter \@secondoftwo
 \fi
}%
\providecommand \natexlab [1]{#1}%
\providecommand \enquote  [1]{``#1''}%
\providecommand \bibnamefont  [1]{#1}%
\providecommand \bibfnamefont [1]{#1}%
\providecommand \citenamefont [1]{#1}%
\providecommand \href@noop [0]{\@secondoftwo}%
\providecommand \href [0]{\begingroup \@sanitize@url \@href}%
\providecommand \@href[1]{\@@startlink{#1}\@@href}%
\providecommand \@@href[1]{\endgroup#1\@@endlink}%
\providecommand \@sanitize@url [0]{\catcode `\\12\catcode `\$12\catcode
  `\&12\catcode `\#12\catcode `\^12\catcode `\_12\catcode `\%12\relax}%
\providecommand \@@startlink[1]{}%
\providecommand \@@endlink[0]{}%
\providecommand \url  [0]{\begingroup\@sanitize@url \@url }%
\providecommand \@url [1]{\endgroup\@href {#1}{\urlprefix }}%
\providecommand \urlprefix  [0]{URL }%
\providecommand \Eprint [0]{\href }%
\providecommand \doibase [0]{http://dx.doi.org/}%
\providecommand \selectlanguage [0]{\@gobble}%
\providecommand \bibinfo  [0]{\@secondoftwo}%
\providecommand \bibfield  [0]{\@secondoftwo}%
\providecommand \translation [1]{[#1]}%
\providecommand \BibitemOpen [0]{}%
\providecommand \bibitemStop [0]{}%
\providecommand \bibitemNoStop [0]{.\EOS\space}%
\providecommand \EOS [0]{\spacefactor3000\relax}%
\providecommand \BibitemShut  [1]{\csname bibitem#1\endcsname}%
\let\auto@bib@innerbib\@empty
\bibitem [{\citenamefont {Kitaev}(2006)}]{kitaev-2006}%
  \BibitemOpen
  \bibfield  {author} {\bibinfo {author} {\bibfnamefont {A.}~\bibnamefont
  {Kitaev}},\ }\bibfield  {title} {\enquote {\bibinfo {title} {Anyons in an
  exactly solved model and beyond},}\ }\href {\doibase
  https://doi.org/10.1016/j.aop.2005.10.005} {\bibfield  {journal} {\bibinfo
  {journal} {Annals of Physics}\ }\textbf {\bibinfo {volume} {321}},\ \bibinfo
  {pages} {2--111} (\bibinfo {year} {2006})},\ \bibinfo {note} {january Special
  Issue}\BibitemShut {NoStop}%
\bibitem [{\citenamefont {Savary}\ and\ \citenamefont
  {Balents}(2016)}]{savary-review}%
  \BibitemOpen
  \bibfield  {author} {\bibinfo {author} {\bibfnamefont {L.}~\bibnamefont
  {Savary}}\ and\ \bibinfo {author} {\bibfnamefont {L.}~\bibnamefont
  {Balents}},\ }\bibfield  {title} {\enquote {\bibinfo {title} {Quantum spin
  liquids: a review},}\ }\href {\doibase 10.1088/0034-4885/80/1/016502}
  {\bibfield  {journal} {\bibinfo  {journal} {Reports on Progress in Physics}\
  }\textbf {\bibinfo {volume} {80}},\ \bibinfo {pages} {016502} (\bibinfo
  {year} {2016})}\BibitemShut {NoStop}%
\bibitem [{\citenamefont {Balents}(2010)}]{Balents2010}%
  \BibitemOpen
  \bibfield  {author} {\bibinfo {author} {\bibfnamefont {L.}~\bibnamefont
  {Balents}},\ }\bibfield  {title} {\enquote {\bibinfo {title} {Spin liquids in
  frustrated magnets},}\ }\href {\doibase 10.1038/nature08917} {\bibfield
  {journal} {\bibinfo  {journal} {Nature}\ }\textbf {\bibinfo {volume} {464}},\
  \bibinfo {pages} {199--208} (\bibinfo {year} {2010})}\BibitemShut {NoStop}%
\bibitem [{\citenamefont {Hermanns}\ \emph {et~al.}(2018)\citenamefont
  {Hermanns}, \citenamefont {Kimchi},\ and\ \citenamefont
  {Knolle}}]{hermanns-2018}%
  \BibitemOpen
  \bibfield  {author} {\bibinfo {author} {\bibfnamefont {M.}~\bibnamefont
  {Hermanns}}, \bibinfo {author} {\bibfnamefont {I.}~\bibnamefont {Kimchi}}, \
  and\ \bibinfo {author} {\bibfnamefont {J.}~\bibnamefont {Knolle}},\
  }\bibfield  {title} {\enquote {\bibinfo {title} {Physics of the kitaev model:
  Fractionalization, dynamic correlations, and material connections},}\ }\href
  {\doibase 10.1146/annurev-conmatphys-033117-053934} {\bibfield  {journal}
  {\bibinfo  {journal} {Annual Review of Condensed Matter Physics}\ }\textbf
  {\bibinfo {volume} {9}},\ \bibinfo {pages} {17--33} (\bibinfo {year}
  {2018})},\ \Eprint
  {http://arxiv.org/abs/https://doi.org/10.1146/annurev-conmatphys-033117-053934}
  {https://doi.org/10.1146/annurev-conmatphys-033117-053934} \BibitemShut
  {NoStop}%
\bibitem [{\citenamefont {Loidl}\ \emph {et~al.}(2021)\citenamefont {Loidl},
  \citenamefont {Lunkenheimer},\ and\ \citenamefont {Tsurkan}}]{Loidl2021}%
  \BibitemOpen
  \bibfield  {author} {\bibinfo {author} {\bibfnamefont {A.}~\bibnamefont
  {Loidl}}, \bibinfo {author} {\bibfnamefont {P.}~\bibnamefont {Lunkenheimer}},
  \ and\ \bibinfo {author} {\bibfnamefont {V.}~\bibnamefont {Tsurkan}},\
  }\bibfield  {title} {\enquote {\bibinfo {title} {On the proximate kitaev
  quantum-spin liquid $\alpha-{\rm rucl_3}$: thermodynamics, excitations and
  continua},}\ }\href {\doibase 10.1088/1361-648X/ac1bcf} {\bibfield  {journal}
  {\bibinfo  {journal} {Journal of Physics: Condensed Matter}\ }\textbf
  {\bibinfo {volume} {33}},\ \bibinfo {pages} {443004} (\bibinfo {year}
  {2021})}\BibitemShut {NoStop}%
\bibitem [{\citenamefont {Trebst}\ and\ \citenamefont
  {Hickey}(2022)}]{simon-2022}%
  \BibitemOpen
  \bibfield  {author} {\bibinfo {author} {\bibfnamefont {S.}~\bibnamefont
  {Trebst}}\ and\ \bibinfo {author} {\bibfnamefont {C.}~\bibnamefont
  {Hickey}},\ }\bibfield  {title} {\enquote {\bibinfo {title} {Kitaev
  materials},}\ }\href {\doibase https://doi.org/10.1016/j.physrep.2021.11.003}
  {\bibfield  {journal} {\bibinfo  {journal} {Physics Reports}\ }\textbf
  {\bibinfo {volume} {950}},\ \bibinfo {pages} {1--37} (\bibinfo {year}
  {2022})},\ \bibinfo {note} {kitaev materials}\BibitemShut {NoStop}%
\bibitem [{\citenamefont {Baskaran}\ \emph {et~al.}(2007)\citenamefont
  {Baskaran}, \citenamefont {Mandal},\ and\ \citenamefont
  {Shankar}}]{smandal-2007}%
  \BibitemOpen
  \bibfield  {author} {\bibinfo {author} {\bibfnamefont {G.}~\bibnamefont
  {Baskaran}}, \bibinfo {author} {\bibfnamefont {S.}~\bibnamefont {Mandal}}, \
  and\ \bibinfo {author} {\bibfnamefont {R.}~\bibnamefont {Shankar}},\
  }\bibfield  {title} {\enquote {\bibinfo {title} {Exact results for spin
  dynamics and fractionalization in the kitaev model},}\ }\href {\doibase
  10.1103/PhysRevLett.98.247201} {\bibfield  {journal} {\bibinfo  {journal}
  {Phys. Rev. Lett.}\ }\textbf {\bibinfo {volume} {98}},\ \bibinfo {pages}
  {247201} (\bibinfo {year} {2007})}\BibitemShut {NoStop}%
\bibitem [{\citenamefont {Sarkar}\ \emph {et~al.}(2020)\citenamefont {Sarkar},
  \citenamefont {Rana},\ and\ \citenamefont {Mandal}}]{subhajit-defect}%
  \BibitemOpen
  \bibfield  {author} {\bibinfo {author} {\bibfnamefont {S.}~\bibnamefont
  {Sarkar}}, \bibinfo {author} {\bibfnamefont {D.}~\bibnamefont {Rana}}, \ and\
  \bibinfo {author} {\bibfnamefont {S.}~\bibnamefont {Mandal}},\ }\bibfield
  {title} {\enquote {\bibinfo {title} {Defect production and quench dynamics in
  the three-dimensional kitaev model},}\ }\href {\doibase
  10.1103/PhysRevB.102.134309} {\bibfield  {journal} {\bibinfo  {journal}
  {Phys. Rev. B}\ }\textbf {\bibinfo {volume} {102}},\ \bibinfo {pages}
  {134309} (\bibinfo {year} {2020})}\BibitemShut {NoStop}%
\bibitem [{\citenamefont {Kao}\ and\ \citenamefont {Perkins}(2021)}]{kao-2021}%
  \BibitemOpen
  \bibfield  {author} {\bibinfo {author} {\bibfnamefont {W.-H.}\ \bibnamefont
  {Kao}}\ and\ \bibinfo {author} {\bibfnamefont {N.~B.}\ \bibnamefont
  {Perkins}},\ }\bibfield  {title} {\enquote {\bibinfo {title} {Disorder upon
  disorder: Localization effects in the kitaev spin liquid},}\ }\href {\doibase
  https://doi.org/10.1016/j.aop.2021.168506} {\bibfield  {journal} {\bibinfo
  {journal} {Annals of Physics}\ }\textbf {\bibinfo {volume} {435}},\ \bibinfo
  {pages} {168506} (\bibinfo {year} {2021})},\ \bibinfo {note} {special issue
  on Philip W. Anderson}\BibitemShut {NoStop}%
\bibitem [{\citenamefont {Nasu}\ and\ \citenamefont
  {Motome}(2020)}]{nasu-2020}%
  \BibitemOpen
  \bibfield  {author} {\bibinfo {author} {\bibfnamefont {J.}~\bibnamefont
  {Nasu}}\ and\ \bibinfo {author} {\bibfnamefont {Y.}~\bibnamefont {Motome}},\
  }\bibfield  {title} {\enquote {\bibinfo {title} {Thermodynamic and transport
  properties in disordered kitaev models},}\ }\href {\doibase
  10.1103/PhysRevB.102.054437} {\bibfield  {journal} {\bibinfo  {journal}
  {Phys. Rev. B}\ }\textbf {\bibinfo {volume} {102}},\ \bibinfo {pages}
  {054437} (\bibinfo {year} {2020})}\BibitemShut {NoStop}%
\bibitem [{\citenamefont {Yao}\ and\ \citenamefont {Qi}(2010)}]{Qi-2010}%
  \BibitemOpen
  \bibfield  {author} {\bibinfo {author} {\bibfnamefont {H.}~\bibnamefont
  {Yao}}\ and\ \bibinfo {author} {\bibfnamefont {X.-L.}\ \bibnamefont {Qi}},\
  }\bibfield  {title} {\enquote {\bibinfo {title} {Entanglement entropy and
  entanglement spectrum of the kitaev model},}\ }\href {\doibase
  10.1103/PhysRevLett.105.080501} {\bibfield  {journal} {\bibinfo  {journal}
  {Phys. Rev. Lett.}\ }\textbf {\bibinfo {volume} {105}},\ \bibinfo {pages}
  {080501} (\bibinfo {year} {2010})}\BibitemShut {NoStop}%
\bibitem [{\citenamefont {Mandal}\ \emph {et~al.}(2016)\citenamefont {Mandal},
  \citenamefont {Maiti},\ and\ \citenamefont {Varma}}]{mandal-2016}%
  \BibitemOpen
  \bibfield  {author} {\bibinfo {author} {\bibfnamefont {S.}~\bibnamefont
  {Mandal}}, \bibinfo {author} {\bibfnamefont {M.}~\bibnamefont {Maiti}}, \
  and\ \bibinfo {author} {\bibfnamefont {V.~K.}\ \bibnamefont {Varma}},\
  }\bibfield  {title} {\enquote {\bibinfo {title} {Entanglement and majorana
  edge states in the kitaev model},}\ }\href {\doibase
  10.1103/PhysRevB.94.045421} {\bibfield  {journal} {\bibinfo  {journal} {Phys.
  Rev. B}\ }\textbf {\bibinfo {volume} {94}},\ \bibinfo {pages} {045421}
  (\bibinfo {year} {2016})}\BibitemShut {NoStop}%
\bibitem [{\citenamefont {Randeep}\ and\ \citenamefont
  {Surendran}(2018)}]{naveen-2018}%
  \BibitemOpen
  \bibfield  {author} {\bibinfo {author} {\bibfnamefont {N.~C.}\ \bibnamefont
  {Randeep}}\ and\ \bibinfo {author} {\bibfnamefont {N.}~\bibnamefont
  {Surendran}},\ }\bibfield  {title} {\enquote {\bibinfo {title} {Topological
  entanglement entropy of the three-dimensional kitaev model},}\ }\href
  {\doibase 10.1103/PhysRevB.98.125136} {\bibfield  {journal} {\bibinfo
  {journal} {Phys. Rev. B}\ }\textbf {\bibinfo {volume} {98}},\ \bibinfo
  {pages} {125136} (\bibinfo {year} {2018})}\BibitemShut {NoStop}%
\bibitem [{\citenamefont {Yao}\ and\ \citenamefont
  {Kivelson}(2007)}]{kivelson-2007}%
  \BibitemOpen
  \bibfield  {author} {\bibinfo {author} {\bibfnamefont {H.}~\bibnamefont
  {Yao}}\ and\ \bibinfo {author} {\bibfnamefont {S.~A.}\ \bibnamefont
  {Kivelson}},\ }\bibfield  {title} {\enquote {\bibinfo {title} {Exact chiral
  spin liquid with non-abelian anyons},}\ }\href {\doibase
  10.1103/PhysRevLett.99.247203} {\bibfield  {journal} {\bibinfo  {journal}
  {Phys. Rev. Lett.}\ }\textbf {\bibinfo {volume} {99}},\ \bibinfo {pages}
  {247203} (\bibinfo {year} {2007})}\BibitemShut {NoStop}%
\bibitem [{\citenamefont {Kells}\ \emph {et~al.}(2010)\citenamefont {Kells},
  \citenamefont {Mehta}, \citenamefont {Slingerland},\ and\ \citenamefont
  {Vala}}]{vala-csl-2010}%
  \BibitemOpen
  \bibfield  {author} {\bibinfo {author} {\bibfnamefont {G.}~\bibnamefont
  {Kells}}, \bibinfo {author} {\bibfnamefont {D.}~\bibnamefont {Mehta}},
  \bibinfo {author} {\bibfnamefont {J.~K.}\ \bibnamefont {Slingerland}}, \ and\
  \bibinfo {author} {\bibfnamefont {J.}~\bibnamefont {Vala}},\ }\bibfield
  {title} {\enquote {\bibinfo {title} {Exact results for the star lattice
  chiral spin liquid},}\ }\href {\doibase 10.1103/PhysRevB.81.104429}
  {\bibfield  {journal} {\bibinfo  {journal} {Phys. Rev. B}\ }\textbf {\bibinfo
  {volume} {81}},\ \bibinfo {pages} {104429} (\bibinfo {year}
  {2010})}\BibitemShut {NoStop}%
\bibitem [{\citenamefont {Mandal}\ and\ \citenamefont
  {Surendran}(2009)}]{naveen-2008}%
  \BibitemOpen
  \bibfield  {author} {\bibinfo {author} {\bibfnamefont {S.}~\bibnamefont
  {Mandal}}\ and\ \bibinfo {author} {\bibfnamefont {N.}~\bibnamefont
  {Surendran}},\ }\bibfield  {title} {\enquote {\bibinfo {title} {Exactly
  solvable kitaev model in three dimensions},}\ }\href {\doibase
  10.1103/PhysRevB.79.024426} {\bibfield  {journal} {\bibinfo  {journal} {Phys.
  Rev. B}\ }\textbf {\bibinfo {volume} {79}},\ \bibinfo {pages} {024426}
  (\bibinfo {year} {2009})}\BibitemShut {NoStop}%
\bibitem [{\citenamefont {Eschmann}\ \emph {et~al.}(2020)\citenamefont
  {Eschmann}, \citenamefont {Mishchenko}, \citenamefont {O'Brien},
  \citenamefont {Bojesen}, \citenamefont {Kato}, \citenamefont {Hermanns},
  \citenamefont {Motome},\ and\ \citenamefont {Trebst}}]{eschmann-2020}%
  \BibitemOpen
  \bibfield  {author} {\bibinfo {author} {\bibfnamefont {T.}~\bibnamefont
  {Eschmann}}, \bibinfo {author} {\bibfnamefont {P.~A.}\ \bibnamefont
  {Mishchenko}}, \bibinfo {author} {\bibfnamefont {K.}~\bibnamefont {O'Brien}},
  \bibinfo {author} {\bibfnamefont {T.~A.}\ \bibnamefont {Bojesen}}, \bibinfo
  {author} {\bibfnamefont {Y.}~\bibnamefont {Kato}}, \bibinfo {author}
  {\bibfnamefont {M.}~\bibnamefont {Hermanns}}, \bibinfo {author}
  {\bibfnamefont {Y.}~\bibnamefont {Motome}}, \ and\ \bibinfo {author}
  {\bibfnamefont {S.}~\bibnamefont {Trebst}},\ }\bibfield  {title} {\enquote
  {\bibinfo {title} {Thermodynamic classification of three-dimensional kitaev
  spin liquids},}\ }\href {\doibase 10.1103/PhysRevB.102.075125} {\bibfield
  {journal} {\bibinfo  {journal} {Phys. Rev. B}\ }\textbf {\bibinfo {volume}
  {102}},\ \bibinfo {pages} {075125} (\bibinfo {year} {2020})}\BibitemShut
  {NoStop}%
\bibitem [{\citenamefont {Mandal}\ and\ \citenamefont
  {Surendran}(2014)}]{mandal-toriccode}%
  \BibitemOpen
  \bibfield  {author} {\bibinfo {author} {\bibfnamefont {S.}~\bibnamefont
  {Mandal}}\ and\ \bibinfo {author} {\bibfnamefont {N.}~\bibnamefont
  {Surendran}},\ }\bibfield  {title} {\enquote {\bibinfo {title} {Fermions and
  nontrivial loop-braiding in a three-dimensional toric code},}\ }\href
  {\doibase 10.1103/PhysRevB.90.104424} {\bibfield  {journal} {\bibinfo
  {journal} {Phys. Rev. B}\ }\textbf {\bibinfo {volume} {90}},\ \bibinfo
  {pages} {104424} (\bibinfo {year} {2014})}\BibitemShut {NoStop}%
\bibitem [{\citenamefont {Mandal}\ \emph
  {et~al.}(2012{\natexlab{a}})\citenamefont {Mandal}, \citenamefont {Shankar},\
  and\ \citenamefont {Baskaran}}]{mandaljpa}%
  \BibitemOpen
  \bibfield  {author} {\bibinfo {author} {\bibfnamefont {S.}~\bibnamefont
  {Mandal}}, \bibinfo {author} {\bibfnamefont {R.}~\bibnamefont {Shankar}}, \
  and\ \bibinfo {author} {\bibfnamefont {G.}~\bibnamefont {Baskaran}},\
  }\bibfield  {title} {\enquote {\bibinfo {title} {Rvb gauge theory and the
  topological degeneracy in the honeycomb kitaev model},}\ }\href {\doibase
  10.1088/1751-8113/45/33/335304} {\bibfield  {journal} {\bibinfo  {journal}
  {Journal of Physics A: Mathematical and Theoretical}\ }\textbf {\bibinfo
  {volume} {45}},\ \bibinfo {pages} {335304} (\bibinfo {year}
  {2012}{\natexlab{a}})}\BibitemShut {NoStop}%
\bibitem [{\citenamefont {Sasidharan}\ and\ \citenamefont
  {Surendran}(2024)}]{Sasidharan_2024}%
  \BibitemOpen
  \bibfield  {author} {\bibinfo {author} {\bibfnamefont {S.}~\bibnamefont
  {Sasidharan}}\ and\ \bibinfo {author} {\bibfnamefont {N.}~\bibnamefont
  {Surendran}},\ }\bibfield  {title} {\enquote {\bibinfo {title} {Periodically
  driven three-dimensional kitaev model},}\ }\href {\doibase
  10.1088/1402-4896/ad3030} {\bibfield  {journal} {\bibinfo  {journal} {Physica
  Scripta}\ }\textbf {\bibinfo {volume} {99}},\ \bibinfo {pages} {045930}
  (\bibinfo {year} {2024})}\BibitemShut {NoStop}%
\bibitem [{\citenamefont {Tikhonov}\ \emph {et~al.}(2011)\citenamefont
  {Tikhonov}, \citenamefont {Feigel'man},\ and\ \citenamefont
  {Kitaev}}]{tikhonov-2011}%
  \BibitemOpen
  \bibfield  {author} {\bibinfo {author} {\bibfnamefont {K.~S.}\ \bibnamefont
  {Tikhonov}}, \bibinfo {author} {\bibfnamefont {M.~V.}\ \bibnamefont
  {Feigel'man}}, \ and\ \bibinfo {author} {\bibfnamefont {A.~Y.}\ \bibnamefont
  {Kitaev}},\ }\bibfield  {title} {\enquote {\bibinfo {title} {Power-law spin
  correlations in a perturbed spin model on a honeycomb lattice},}\ }\href
  {\doibase 10.1103/PhysRevLett.106.067203} {\bibfield  {journal} {\bibinfo
  {journal} {Phys. Rev. Lett.}\ }\textbf {\bibinfo {volume} {106}},\ \bibinfo
  {pages} {067203} (\bibinfo {year} {2011})}\BibitemShut {NoStop}%
\bibitem [{\citenamefont {Lunkin}\ \emph {et~al.}(2019)\citenamefont {Lunkin},
  \citenamefont {Tikhonov},\ and\ \citenamefont {Feigel'man}}]{lunkin-2019}%
  \BibitemOpen
  \bibfield  {author} {\bibinfo {author} {\bibfnamefont {A.}~\bibnamefont
  {Lunkin}}, \bibinfo {author} {\bibfnamefont {K.}~\bibnamefont {Tikhonov}}, \
  and\ \bibinfo {author} {\bibfnamefont {M.}~\bibnamefont {Feigel'man}},\
  }\bibfield  {title} {\enquote {\bibinfo {title} {Perturbed kitaev model:
  Excitation spectrum and long-ranged spin correlations},}\ }\href {\doibase
  https://doi.org/10.1016/j.jpcs.2017.11.009} {\bibfield  {journal} {\bibinfo
  {journal} {Journal of Physics and Chemistry of Solids}\ }\textbf {\bibinfo
  {volume} {128}},\ \bibinfo {pages} {130--137} (\bibinfo {year} {2019})},\
  \bibinfo {note} {spin-Orbit Coupled Materials}\BibitemShut {NoStop}%
\bibitem [{\citenamefont {Mandal}\ \emph {et~al.}(2011)\citenamefont {Mandal},
  \citenamefont {Bhattacharjee}, \citenamefont {Sengupta}, \citenamefont
  {Shankar},\ and\ \citenamefont {Baskaran}}]{mandal-subhro-2011}%
  \BibitemOpen
  \bibfield  {author} {\bibinfo {author} {\bibfnamefont {S.}~\bibnamefont
  {Mandal}}, \bibinfo {author} {\bibfnamefont {S.}~\bibnamefont
  {Bhattacharjee}}, \bibinfo {author} {\bibfnamefont {K.}~\bibnamefont
  {Sengupta}}, \bibinfo {author} {\bibfnamefont {R.}~\bibnamefont {Shankar}}, \
  and\ \bibinfo {author} {\bibfnamefont {G.}~\bibnamefont {Baskaran}},\
  }\bibfield  {title} {\enquote {\bibinfo {title} {Confinement-deconfinement
  transition and spin correlations in a generalized kitaev model},}\ }\href
  {\doibase 10.1103/PhysRevB.84.155121} {\bibfield  {journal} {\bibinfo
  {journal} {Phys. Rev. B}\ }\textbf {\bibinfo {volume} {84}},\ \bibinfo
  {pages} {155121} (\bibinfo {year} {2011})}\BibitemShut {NoStop}%
\bibitem [{\citenamefont {Chaloupka}\ \emph {et~al.}(2010)\citenamefont
  {Chaloupka}, \citenamefont {Jackeli},\ and\ \citenamefont
  {Khaliullin}}]{chalaupka-2010}%
  \BibitemOpen
  \bibfield  {author} {\bibinfo {author} {\bibfnamefont {J.~c.~v.}\
  \bibnamefont {Chaloupka}}, \bibinfo {author} {\bibfnamefont {G.}~\bibnamefont
  {Jackeli}}, \ and\ \bibinfo {author} {\bibfnamefont {G.}~\bibnamefont
  {Khaliullin}},\ }\bibfield  {title} {\enquote {\bibinfo {title}
  {Kitaev-heisenberg model on a honeycomb lattice: Possible exotic phases in
  iridium oxides ${A}_{2}{\mathrm{iro}}_{3}$},}\ }\href {\doibase
  10.1103/PhysRevLett.105.027204} {\bibfield  {journal} {\bibinfo  {journal}
  {Phys. Rev. Lett.}\ }\textbf {\bibinfo {volume} {105}},\ \bibinfo {pages}
  {027204} (\bibinfo {year} {2010})}\BibitemShut {NoStop}%
\bibitem [{\citenamefont {Nanda}\ \emph {et~al.}(2020)\citenamefont {Nanda},
  \citenamefont {Dhochak},\ and\ \citenamefont {Bhattacharjee}}]{animesh-2020}%
  \BibitemOpen
  \bibfield  {author} {\bibinfo {author} {\bibfnamefont {A.}~\bibnamefont
  {Nanda}}, \bibinfo {author} {\bibfnamefont {K.}~\bibnamefont {Dhochak}}, \
  and\ \bibinfo {author} {\bibfnamefont {S.}~\bibnamefont {Bhattacharjee}},\
  }\bibfield  {title} {\enquote {\bibinfo {title} {Phases and quantum phase
  transitions in an anisotropic ferromagnetic
  kitaev-heisenberg-$\mathrm{\ensuremath{\Gamma}}$ magnet},}\ }\href {\doibase
  10.1103/PhysRevB.102.235124} {\bibfield  {journal} {\bibinfo  {journal}
  {Phys. Rev. B}\ }\textbf {\bibinfo {volume} {102}},\ \bibinfo {pages}
  {235124} (\bibinfo {year} {2020})}\BibitemShut {NoStop}%
\bibitem [{\citenamefont {Nanda}\ \emph {et~al.}(2021)\citenamefont {Nanda},
  \citenamefont {Agarwala},\ and\ \citenamefont
  {Bhattacharjee}}]{animesh-2021}%
  \BibitemOpen
  \bibfield  {author} {\bibinfo {author} {\bibfnamefont {A.}~\bibnamefont
  {Nanda}}, \bibinfo {author} {\bibfnamefont {A.}~\bibnamefont {Agarwala}}, \
  and\ \bibinfo {author} {\bibfnamefont {S.}~\bibnamefont {Bhattacharjee}},\
  }\bibfield  {title} {\enquote {\bibinfo {title} {Phases and quantum phase
  transitions in the anisotropic antiferromagnetic
  kitaev-heisenberg-$\mathrm{\ensuremath{\Gamma}}$ magnet},}\ }\href {\doibase
  10.1103/PhysRevB.104.195115} {\bibfield  {journal} {\bibinfo  {journal}
  {Phys. Rev. B}\ }\textbf {\bibinfo {volume} {104}},\ \bibinfo {pages}
  {195115} (\bibinfo {year} {2021})}\BibitemShut {NoStop}%
\bibitem [{\citenamefont {Knolle}\ \emph {et~al.}(2018)\citenamefont {Knolle},
  \citenamefont {Bhattacharjee},\ and\ \citenamefont {Moessner}}]{knolle-2018}%
  \BibitemOpen
  \bibfield  {author} {\bibinfo {author} {\bibfnamefont {J.}~\bibnamefont
  {Knolle}}, \bibinfo {author} {\bibfnamefont {S.}~\bibnamefont
  {Bhattacharjee}}, \ and\ \bibinfo {author} {\bibfnamefont {R.}~\bibnamefont
  {Moessner}},\ }\bibfield  {title} {\enquote {\bibinfo {title} {Dynamics of a
  quantum spin liquid beyond integrability: The
  kitaev-heisenberg-$\mathrm{\ensuremath{\Gamma}}$ model in an augmented parton
  mean-field theory},}\ }\href {\doibase 10.1103/PhysRevB.97.134432} {\bibfield
   {journal} {\bibinfo  {journal} {Phys. Rev. B}\ }\textbf {\bibinfo {volume}
  {97}},\ \bibinfo {pages} {134432} (\bibinfo {year} {2018})}\BibitemShut
  {NoStop}%
\bibitem [{\citenamefont {Rau}\ \emph {et~al.}(2014)\citenamefont {Rau},
  \citenamefont {Lee},\ and\ \citenamefont {Kee}}]{Rau-2014}%
  \BibitemOpen
  \bibfield  {author} {\bibinfo {author} {\bibfnamefont {J.~G.}\ \bibnamefont
  {Rau}}, \bibinfo {author} {\bibfnamefont {E.~K.-H.}\ \bibnamefont {Lee}}, \
  and\ \bibinfo {author} {\bibfnamefont {H.-Y.}\ \bibnamefont {Kee}},\
  }\bibfield  {title} {\enquote {\bibinfo {title} {Generic spin model for the
  honeycomb iridates beyond the kitaev limit},}\ }\href {\doibase
  10.1103/PhysRevLett.112.077204} {\bibfield  {journal} {\bibinfo  {journal}
  {Phys. Rev. Lett.}\ }\textbf {\bibinfo {volume} {112}},\ \bibinfo {pages}
  {077204} (\bibinfo {year} {2014})}\BibitemShut {NoStop}%
\bibitem [{\citenamefont {Banerjee}\ and\ \citenamefont
  {Lin}(2023)}]{saikat-lin}%
  \BibitemOpen
  \bibfield  {author} {\bibinfo {author} {\bibfnamefont {S.}~\bibnamefont
  {Banerjee}}\ and\ \bibinfo {author} {\bibfnamefont {S.-Z.}\ \bibnamefont
  {Lin}},\ }\bibfield  {title} {\enquote {\bibinfo {title} {{Emergent orbital
  magnetization in Kitaev quantum magnets}},}\ }\href {\doibase
  10.21468/SciPostPhys.14.5.127} {\bibfield  {journal} {\bibinfo  {journal}
  {SciPost Phys.}\ }\textbf {\bibinfo {volume} {14}},\ \bibinfo {pages} {127}
  (\bibinfo {year} {2023})}\BibitemShut {NoStop}%
\bibitem [{\citenamefont {Kumar}\ \emph {et~al.}(2022)\citenamefont {Kumar},
  \citenamefont {Banerjee},\ and\ \citenamefont {Lin}}]{Kumar2022}%
  \BibitemOpen
  \bibfield  {author} {\bibinfo {author} {\bibfnamefont {U.}~\bibnamefont
  {Kumar}}, \bibinfo {author} {\bibfnamefont {S.}~\bibnamefont {Banerjee}}, \
  and\ \bibinfo {author} {\bibfnamefont {S.-Z.}\ \bibnamefont {Lin}},\
  }\bibfield  {title} {\enquote {\bibinfo {title} {Floquet engineering of
  kitaev quantum magnets},}\ }\href {\doibase 10.1038/s42005-022-00931-1}
  {\bibfield  {journal} {\bibinfo  {journal} {Communications Physics}\ }\textbf
  {\bibinfo {volume} {5}},\ \bibinfo {pages} {157} (\bibinfo {year}
  {2022})}\BibitemShut {NoStop}%
\bibitem [{\citenamefont {Banerjee}\ \emph {et~al.}(2022)\citenamefont
  {Banerjee}, \citenamefont {Kumar},\ and\ \citenamefont
  {Lin}}]{saikat-inverseFARADAY}%
  \BibitemOpen
  \bibfield  {author} {\bibinfo {author} {\bibfnamefont {S.}~\bibnamefont
  {Banerjee}}, \bibinfo {author} {\bibfnamefont {U.}~\bibnamefont {Kumar}}, \
  and\ \bibinfo {author} {\bibfnamefont {S.-Z.}\ \bibnamefont {Lin}},\
  }\bibfield  {title} {\enquote {\bibinfo {title} {Inverse faraday effect in
  mott insulators},}\ }\href {\doibase 10.1103/PhysRevB.105.L180414} {\bibfield
   {journal} {\bibinfo  {journal} {Phys. Rev. B}\ }\textbf {\bibinfo {volume}
  {105}},\ \bibinfo {pages} {L180414} (\bibinfo {year} {2022})}\BibitemShut
  {NoStop}%
\bibitem [{\citenamefont {Takagi}\ \emph {et~al.}(2019)\citenamefont {Takagi},
  \citenamefont {Takayama}, \citenamefont {Jackeli}, \citenamefont
  {Khaliullin},\ and\ \citenamefont {Nagler}}]{takagi-2019}%
  \BibitemOpen
  \bibfield  {author} {\bibinfo {author} {\bibfnamefont {H.}~\bibnamefont
  {Takagi}}, \bibinfo {author} {\bibfnamefont {T.}~\bibnamefont {Takayama}},
  \bibinfo {author} {\bibfnamefont {G.}~\bibnamefont {Jackeli}}, \bibinfo
  {author} {\bibfnamefont {G.}~\bibnamefont {Khaliullin}}, \ and\ \bibinfo
  {author} {\bibfnamefont {S.~E.}\ \bibnamefont {Nagler}},\ }\bibfield  {title}
  {\enquote {\bibinfo {title} {Concept and realization of kitaev quantum spin
  liquids},}\ }\href {\doibase 10.1038/s42254-019-0038-2} {\bibfield  {journal}
  {\bibinfo  {journal} {Nature Reviews Physics}\ }\textbf {\bibinfo {volume}
  {1}},\ \bibinfo {pages} {264--280} (\bibinfo {year} {2019})}\BibitemShut
  {NoStop}%
\bibitem [{\citenamefont {Motome}\ \emph {et~al.}(2020)\citenamefont {Motome},
  \citenamefont {Sano}, \citenamefont {Jang}, \citenamefont {Sugita},\ and\
  \citenamefont {Kato}}]{motome-2020}%
  \BibitemOpen
  \bibfield  {author} {\bibinfo {author} {\bibfnamefont {Y.}~\bibnamefont
  {Motome}}, \bibinfo {author} {\bibfnamefont {R.}~\bibnamefont {Sano}},
  \bibinfo {author} {\bibfnamefont {S.}~\bibnamefont {Jang}}, \bibinfo {author}
  {\bibfnamefont {Y.}~\bibnamefont {Sugita}}, \ and\ \bibinfo {author}
  {\bibfnamefont {Y.}~\bibnamefont {Kato}},\ }\bibfield  {title} {\enquote
  {\bibinfo {title} {Materials design of kitaev spin liquids beyond the
  jackeli–khaliullin mechanism},}\ }\href {\doibase 10.1088/1361-648X/ab8525}
  {\bibfield  {journal} {\bibinfo  {journal} {Journal of Physics: Condensed
  Matter}\ }\textbf {\bibinfo {volume} {32}},\ \bibinfo {pages} {404001}
  (\bibinfo {year} {2020})}\BibitemShut {NoStop}%
\bibitem [{\citenamefont {Banerjee}\ \emph {et~al.}(2016)\citenamefont
  {Banerjee}, \citenamefont {Bridges}, \citenamefont {Yan}, \citenamefont
  {Aczel}, \citenamefont {Li}, \citenamefont {Stone}, \citenamefont {Granroth},
  \citenamefont {Lumsden}, \citenamefont {Yiu}, \citenamefont {Knolle},
  \citenamefont {Bhattacharjee}, \citenamefont {Kovrizhin}, \citenamefont
  {Moessner}, \citenamefont {Tennant}, \citenamefont {Mandrus},\ and\
  \citenamefont {Nagler}}]{abanerjee-2016}%
  \BibitemOpen
  \bibfield  {author} {\bibinfo {author} {\bibfnamefont {A.}~\bibnamefont
  {Banerjee}}, \bibinfo {author} {\bibfnamefont {C.~A.}\ \bibnamefont
  {Bridges}}, \bibinfo {author} {\bibfnamefont {J.-Q.}\ \bibnamefont {Yan}},
  \bibinfo {author} {\bibfnamefont {A.~A.}\ \bibnamefont {Aczel}}, \bibinfo
  {author} {\bibfnamefont {L.}~\bibnamefont {Li}}, \bibinfo {author}
  {\bibfnamefont {M.~B.}\ \bibnamefont {Stone}}, \bibinfo {author}
  {\bibfnamefont {G.~E.}\ \bibnamefont {Granroth}}, \bibinfo {author}
  {\bibfnamefont {M.~D.}\ \bibnamefont {Lumsden}}, \bibinfo {author}
  {\bibfnamefont {Y.}~\bibnamefont {Yiu}}, \bibinfo {author} {\bibfnamefont
  {J.}~\bibnamefont {Knolle}}, \bibinfo {author} {\bibfnamefont
  {S.}~\bibnamefont {Bhattacharjee}}, \bibinfo {author} {\bibfnamefont {D.~L.}\
  \bibnamefont {Kovrizhin}}, \bibinfo {author} {\bibfnamefont {R.}~\bibnamefont
  {Moessner}}, \bibinfo {author} {\bibfnamefont {D.~A.}\ \bibnamefont
  {Tennant}}, \bibinfo {author} {\bibfnamefont {D.~G.}\ \bibnamefont
  {Mandrus}}, \ and\ \bibinfo {author} {\bibfnamefont {S.~E.}\ \bibnamefont
  {Nagler}},\ }\bibfield  {title} {\enquote {\bibinfo {title} {Proximate kitaev
  quantum spin liquid behaviour in a honeycomb magnet},}\ }\href {\doibase
  10.1038/nmat4604} {\bibfield  {journal} {\bibinfo  {journal} {Nature
  Materials}\ }\textbf {\bibinfo {volume} {15}},\ \bibinfo {pages} {733--740}
  (\bibinfo {year} {2016})}\BibitemShut {NoStop}%
\bibitem [{\citenamefont {Janssen}\ \emph {et~al.}(2017)\citenamefont
  {Janssen}, \citenamefont {Andrade},\ and\ \citenamefont
  {Vojta}}]{janssen-2017}%
  \BibitemOpen
  \bibfield  {author} {\bibinfo {author} {\bibfnamefont {L.}~\bibnamefont
  {Janssen}}, \bibinfo {author} {\bibfnamefont {E.~C.}\ \bibnamefont
  {Andrade}}, \ and\ \bibinfo {author} {\bibfnamefont {M.}~\bibnamefont
  {Vojta}},\ }\bibfield  {title} {\enquote {\bibinfo {title} {Magnetization
  processes of zigzag states on the honeycomb lattice: Identifying spin models
  for $\ensuremath{\alpha}\text{\ensuremath{-}}{\mathrm{rucl}}_{3}$ and
  ${\mathrm{na}}_{2}{\mathrm{iro}}_{3}$},}\ }\href {\doibase
  10.1103/PhysRevB.96.064430} {\bibfield  {journal} {\bibinfo  {journal} {Phys.
  Rev. B}\ }\textbf {\bibinfo {volume} {96}},\ \bibinfo {pages} {064430}
  (\bibinfo {year} {2017})}\BibitemShut {NoStop}%
\bibitem [{\citenamefont {Chern}\ \emph {et~al.}(2021)\citenamefont {Chern},
  \citenamefont {Buessen},\ and\ \citenamefont {Kim}}]{chern-2021}%
  \BibitemOpen
  \bibfield  {author} {\bibinfo {author} {\bibfnamefont {L.~E.}\ \bibnamefont
  {Chern}}, \bibinfo {author} {\bibfnamefont {F.~L.}\ \bibnamefont {Buessen}},
  \ and\ \bibinfo {author} {\bibfnamefont {Y.~B.}\ \bibnamefont {Kim}},\
  }\bibfield  {title} {\enquote {\bibinfo {title} {Classical magnetic vortex
  liquid and large thermal hall conductivity in frustrated magnets with
  bond-dependent interactions},}\ }\href {\doibase 10.1038/s41535-021-00331-8}
  {\bibfield  {journal} {\bibinfo  {journal} {npj Quantum Materials}\ }\textbf
  {\bibinfo {volume} {6}},\ \bibinfo {pages} {33} (\bibinfo {year}
  {2021})}\BibitemShut {NoStop}%
\bibitem [{\citenamefont {Czajka}\ \emph {et~al.}(2021)\citenamefont {Czajka},
  \citenamefont {Gao}, \citenamefont {Hirschberger}, \citenamefont
  {Lampen-Kelley}, \citenamefont {Banerjee}, \citenamefont {Yan}, \citenamefont
  {Mandrus}, \citenamefont {Nagler},\ and\ \citenamefont {Ong}}]{czajka-2021}%
  \BibitemOpen
  \bibfield  {author} {\bibinfo {author} {\bibfnamefont {P.}~\bibnamefont
  {Czajka}}, \bibinfo {author} {\bibfnamefont {T.}~\bibnamefont {Gao}},
  \bibinfo {author} {\bibfnamefont {M.}~\bibnamefont {Hirschberger}}, \bibinfo
  {author} {\bibfnamefont {P.}~\bibnamefont {Lampen-Kelley}}, \bibinfo {author}
  {\bibfnamefont {A.}~\bibnamefont {Banerjee}}, \bibinfo {author}
  {\bibfnamefont {J.}~\bibnamefont {Yan}}, \bibinfo {author} {\bibfnamefont
  {D.~G.}\ \bibnamefont {Mandrus}}, \bibinfo {author} {\bibfnamefont {S.~E.}\
  \bibnamefont {Nagler}}, \ and\ \bibinfo {author} {\bibfnamefont {N.~P.}\
  \bibnamefont {Ong}},\ }\bibfield  {title} {\enquote {\bibinfo {title}
  {Oscillations of the thermal conductivity in the spin-liquid state of
  $\alpha$-rucl3},}\ }\href {\doibase 10.1038/s41567-021-01243-x} {\bibfield
  {journal} {\bibinfo  {journal} {Nature Physics}\ }\textbf {\bibinfo {volume}
  {17}},\ \bibinfo {pages} {915--919} (\bibinfo {year} {2021})}\BibitemShut
  {NoStop}%
\bibitem [{\citenamefont {Patel}\ and\ \citenamefont
  {Trivedi}(2019)}]{niravkumar-2019}%
  \BibitemOpen
  \bibfield  {author} {\bibinfo {author} {\bibfnamefont {N.~D.}\ \bibnamefont
  {Patel}}\ and\ \bibinfo {author} {\bibfnamefont {N.}~\bibnamefont
  {Trivedi}},\ }\bibfield  {title} {\enquote {\bibinfo {title} {Magnetic
  field-induced intermediate quantum spin liquid with a spinon fermi
  surface},}\ }\href {\doibase 10.1073/pnas.1821406116} {\bibfield  {journal}
  {\bibinfo  {journal} {Proceedings of the National Academy of Sciences}\
  }\textbf {\bibinfo {volume} {116}},\ \bibinfo {pages} {12199--12203}
  (\bibinfo {year} {2019})},\ \Eprint
  {http://arxiv.org/abs/https://www.pnas.org/doi/pdf/10.1073/pnas.1821406116}
  {https://www.pnas.org/doi/pdf/10.1073/pnas.1821406116} \BibitemShut {NoStop}%
\bibitem [{\citenamefont {Wulferding}\ \emph {et~al.}(2020)\citenamefont
  {Wulferding}, \citenamefont {Choi}, \citenamefont {Do}, \citenamefont {Lee},
  \citenamefont {Lemmens}, \citenamefont {Faugeras}, \citenamefont {Gallais},\
  and\ \citenamefont {Choi}}]{wulferding-2020}%
  \BibitemOpen
  \bibfield  {author} {\bibinfo {author} {\bibfnamefont {D.}~\bibnamefont
  {Wulferding}}, \bibinfo {author} {\bibfnamefont {Y.}~\bibnamefont {Choi}},
  \bibinfo {author} {\bibfnamefont {S.-H.}\ \bibnamefont {Do}}, \bibinfo
  {author} {\bibfnamefont {C.~H.}\ \bibnamefont {Lee}}, \bibinfo {author}
  {\bibfnamefont {P.}~\bibnamefont {Lemmens}}, \bibinfo {author} {\bibfnamefont
  {C.}~\bibnamefont {Faugeras}}, \bibinfo {author} {\bibfnamefont
  {Y.}~\bibnamefont {Gallais}}, \ and\ \bibinfo {author} {\bibfnamefont
  {K.-Y.}\ \bibnamefont {Choi}},\ }\bibfield  {title} {\enquote {\bibinfo
  {title} {Magnon bound states versus anyonic majorana excitations in the
  kitaev honeycomb magnet $\alpha$-rucl3},}\ }\href {\doibase
  10.1038/s41467-020-15370-1} {\bibfield  {journal} {\bibinfo  {journal}
  {Nature Communications}\ }\textbf {\bibinfo {volume} {11}},\ \bibinfo {pages}
  {1603} (\bibinfo {year} {2020})}\BibitemShut {NoStop}%
\bibitem [{\citenamefont {Berke}\ \emph {et~al.}(2020)\citenamefont {Berke},
  \citenamefont {Trebst},\ and\ \citenamefont {Hickey}}]{berke-prb-2020}%
  \BibitemOpen
  \bibfield  {author} {\bibinfo {author} {\bibfnamefont {C.}~\bibnamefont
  {Berke}}, \bibinfo {author} {\bibfnamefont {S.}~\bibnamefont {Trebst}}, \
  and\ \bibinfo {author} {\bibfnamefont {C.}~\bibnamefont {Hickey}},\
  }\bibfield  {title} {\enquote {\bibinfo {title} {Field stability of majorana
  spin liquids in antiferromagnetic kitaev models},}\ }\href {\doibase
  10.1103/PhysRevB.101.214442} {\bibfield  {journal} {\bibinfo  {journal}
  {Phys. Rev. B}\ }\textbf {\bibinfo {volume} {101}},\ \bibinfo {pages}
  {214442} (\bibinfo {year} {2020})}\BibitemShut {NoStop}%
\bibitem [{\citenamefont {Li}\ \emph {et~al.}(2021{\natexlab{a}})\citenamefont
  {Li}, \citenamefont {Zhang}, \citenamefont {Said}, \citenamefont {Fabbris},
  \citenamefont {Mazzone}, \citenamefont {Yan}, \citenamefont {Mandrus},
  \citenamefont {Hal{\'a}sz}, \citenamefont {Okamoto}, \citenamefont
  {Murakami}, \citenamefont {Dean}, \citenamefont {Lee},\ and\ \citenamefont
  {Miao}}]{haoxiang-nature-2012}%
  \BibitemOpen
  \bibfield  {author} {\bibinfo {author} {\bibfnamefont {H.}~\bibnamefont
  {Li}}, \bibinfo {author} {\bibfnamefont {T.~T.}\ \bibnamefont {Zhang}},
  \bibinfo {author} {\bibfnamefont {A.}~\bibnamefont {Said}}, \bibinfo {author}
  {\bibfnamefont {G.}~\bibnamefont {Fabbris}}, \bibinfo {author} {\bibfnamefont
  {D.~G.}\ \bibnamefont {Mazzone}}, \bibinfo {author} {\bibfnamefont {J.~Q.}\
  \bibnamefont {Yan}}, \bibinfo {author} {\bibfnamefont {D.}~\bibnamefont
  {Mandrus}}, \bibinfo {author} {\bibfnamefont {G.~B.}\ \bibnamefont
  {Hal{\'a}sz}}, \bibinfo {author} {\bibfnamefont {S.}~\bibnamefont {Okamoto}},
  \bibinfo {author} {\bibfnamefont {S.}~\bibnamefont {Murakami}}, \bibinfo
  {author} {\bibfnamefont {M.~P.~M.}\ \bibnamefont {Dean}}, \bibinfo {author}
  {\bibfnamefont {H.~N.}\ \bibnamefont {Lee}}, \ and\ \bibinfo {author}
  {\bibfnamefont {H.}~\bibnamefont {Miao}},\ }\bibfield  {title} {\enquote
  {\bibinfo {title} {Giant phonon anomalies in the proximate kitaev quantum
  spin liquid $\alpha$-rucl3},}\ }\href {\doibase 10.1038/s41467-021-23826-1}
  {\bibfield  {journal} {\bibinfo  {journal} {Nature Communications}\ }\textbf
  {\bibinfo {volume} {12}},\ \bibinfo {pages} {3513} (\bibinfo {year}
  {2021}{\natexlab{a}})}\BibitemShut {NoStop}%
\bibitem [{\citenamefont {Balz}\ \emph {et~al.}(2021)\citenamefont {Balz},
  \citenamefont {Janssen}, \citenamefont {Lampen-Kelley}, \citenamefont
  {Banerjee}, \citenamefont {Liu}, \citenamefont {Yan}, \citenamefont
  {Mandrus}, \citenamefont {Vojta},\ and\ \citenamefont
  {Nagler}}]{balz-prb-2021}%
  \BibitemOpen
  \bibfield  {author} {\bibinfo {author} {\bibfnamefont {C.}~\bibnamefont
  {Balz}}, \bibinfo {author} {\bibfnamefont {L.}~\bibnamefont {Janssen}},
  \bibinfo {author} {\bibfnamefont {P.}~\bibnamefont {Lampen-Kelley}}, \bibinfo
  {author} {\bibfnamefont {A.}~\bibnamefont {Banerjee}}, \bibinfo {author}
  {\bibfnamefont {Y.~H.}\ \bibnamefont {Liu}}, \bibinfo {author} {\bibfnamefont
  {J.-Q.}\ \bibnamefont {Yan}}, \bibinfo {author} {\bibfnamefont {D.~G.}\
  \bibnamefont {Mandrus}}, \bibinfo {author} {\bibfnamefont {M.}~\bibnamefont
  {Vojta}}, \ and\ \bibinfo {author} {\bibfnamefont {S.~E.}\ \bibnamefont
  {Nagler}},\ }\bibfield  {title} {\enquote {\bibinfo {title} {Field-induced
  intermediate ordered phase and anisotropic interlayer interactions in
  $\ensuremath{\alpha}\text{\ensuremath{-}}{\mathrm{rucl}}_{3}$},}\ }\href
  {\doibase 10.1103/PhysRevB.103.174417} {\bibfield  {journal} {\bibinfo
  {journal} {Phys. Rev. B}\ }\textbf {\bibinfo {volume} {103}},\ \bibinfo
  {pages} {174417} (\bibinfo {year} {2021})}\BibitemShut {NoStop}%
\bibitem [{\citenamefont {Reig-i Plessis}\ \emph {et~al.}(2020)\citenamefont
  {Reig-i Plessis}, \citenamefont {Johnson}, \citenamefont {Lu}, \citenamefont
  {Chen}, \citenamefont {Ruff}, \citenamefont {Upton}, \citenamefont
  {Williams}, \citenamefont {Calder}, \citenamefont {Zhou}, \citenamefont
  {Clancy}, \citenamefont {Aczel},\ and\ \citenamefont
  {MacDougall}}]{plessis-2020}%
  \BibitemOpen
  \bibfield  {author} {\bibinfo {author} {\bibfnamefont {D.}~\bibnamefont
  {Reig-i Plessis}}, \bibinfo {author} {\bibfnamefont {T.~A.}\ \bibnamefont
  {Johnson}}, \bibinfo {author} {\bibfnamefont {K.}~\bibnamefont {Lu}},
  \bibinfo {author} {\bibfnamefont {Q.}~\bibnamefont {Chen}}, \bibinfo {author}
  {\bibfnamefont {J.~P.~C.}\ \bibnamefont {Ruff}}, \bibinfo {author}
  {\bibfnamefont {M.~H.}\ \bibnamefont {Upton}}, \bibinfo {author}
  {\bibfnamefont {T.~J.}\ \bibnamefont {Williams}}, \bibinfo {author}
  {\bibfnamefont {S.}~\bibnamefont {Calder}}, \bibinfo {author} {\bibfnamefont
  {H.~D.}\ \bibnamefont {Zhou}}, \bibinfo {author} {\bibfnamefont {J.~P.}\
  \bibnamefont {Clancy}}, \bibinfo {author} {\bibfnamefont {A.~A.}\
  \bibnamefont {Aczel}}, \ and\ \bibinfo {author} {\bibfnamefont {G.~J.}\
  \bibnamefont {MacDougall}},\ }\bibfield  {title} {\enquote {\bibinfo {title}
  {Structural, electronic, and magnetic properties of nearly ideal
  ${J}_{eff}=\frac{1}{2}$ iridium halides},}\ }\href {\doibase
  10.1103/PhysRevMaterials.4.124407} {\bibfield  {journal} {\bibinfo  {journal}
  {Phys. Rev. Mater.}\ }\textbf {\bibinfo {volume} {4}},\ \bibinfo {pages}
  {124407} (\bibinfo {year} {2020})}\BibitemShut {NoStop}%
\bibitem [{\citenamefont {Biswas}\ \emph {et~al.}(2019)\citenamefont {Biswas},
  \citenamefont {Li}, \citenamefont {Winter}, \citenamefont {Knolle},\ and\
  \citenamefont {Valent\'{\i}}}]{sananda-2019}%
  \BibitemOpen
  \bibfield  {author} {\bibinfo {author} {\bibfnamefont {S.}~\bibnamefont
  {Biswas}}, \bibinfo {author} {\bibfnamefont {Y.}~\bibnamefont {Li}}, \bibinfo
  {author} {\bibfnamefont {S.~M.}\ \bibnamefont {Winter}}, \bibinfo {author}
  {\bibfnamefont {J.}~\bibnamefont {Knolle}}, \ and\ \bibinfo {author}
  {\bibfnamefont {R.}~\bibnamefont {Valent\'{\i}}},\ }\bibfield  {title}
  {\enquote {\bibinfo {title} {Electronic properties of
  $\ensuremath{\alpha}\text{\ensuremath{-}}{\mathrm{rucl}}_{3}$ in proximity to
  graphene},}\ }\href {\doibase 10.1103/PhysRevLett.123.237201} {\bibfield
  {journal} {\bibinfo  {journal} {Phys. Rev. Lett.}\ }\textbf {\bibinfo
  {volume} {123}},\ \bibinfo {pages} {237201} (\bibinfo {year}
  {2019})}\BibitemShut {NoStop}%
\bibitem [{\citenamefont {Leeb}\ \emph {et~al.}(2021)\citenamefont {Leeb},
  \citenamefont {Polyudov}, \citenamefont {Mashhadi}, \citenamefont {Biswas},
  \citenamefont {Valent\'{\i}}, \citenamefont {Burghard},\ and\ \citenamefont
  {Knolle}}]{sananda-2021}%
  \BibitemOpen
  \bibfield  {author} {\bibinfo {author} {\bibfnamefont {V.}~\bibnamefont
  {Leeb}}, \bibinfo {author} {\bibfnamefont {K.}~\bibnamefont {Polyudov}},
  \bibinfo {author} {\bibfnamefont {S.}~\bibnamefont {Mashhadi}}, \bibinfo
  {author} {\bibfnamefont {S.}~\bibnamefont {Biswas}}, \bibinfo {author}
  {\bibfnamefont {R.}~\bibnamefont {Valent\'{\i}}}, \bibinfo {author}
  {\bibfnamefont {M.}~\bibnamefont {Burghard}}, \ and\ \bibinfo {author}
  {\bibfnamefont {J.}~\bibnamefont {Knolle}},\ }\bibfield  {title} {\enquote
  {\bibinfo {title} {Anomalous quantum oscillations in a heterostructure of
  graphene on a proximate quantum spin liquid},}\ }\href {\doibase
  10.1103/PhysRevLett.126.097201} {\bibfield  {journal} {\bibinfo  {journal}
  {Phys. Rev. Lett.}\ }\textbf {\bibinfo {volume} {126}},\ \bibinfo {pages}
  {097201} (\bibinfo {year} {2021})}\BibitemShut {NoStop}%
\bibitem [{\citenamefont {Nasu}\ and\ \citenamefont
  {Motome}(2019)}]{nasu-motome-2019}%
  \BibitemOpen
  \bibfield  {author} {\bibinfo {author} {\bibfnamefont {J.}~\bibnamefont
  {Nasu}}\ and\ \bibinfo {author} {\bibfnamefont {Y.}~\bibnamefont {Motome}},\
  }\bibfield  {title} {\enquote {\bibinfo {title} {Nonequilibrium majorana
  dynamics by quenching a magnetic field in kitaev spin liquids},}\ }\href
  {\doibase 10.1103/PhysRevResearch.1.033007} {\bibfield  {journal} {\bibinfo
  {journal} {Phys. Rev. Res.}\ }\textbf {\bibinfo {volume} {1}},\ \bibinfo
  {pages} {033007} (\bibinfo {year} {2019})}\BibitemShut {NoStop}%
\bibitem [{\citenamefont {Ronquillo}\ \emph {et~al.}(2019)\citenamefont
  {Ronquillo}, \citenamefont {Vengal},\ and\ \citenamefont
  {Trivedi}}]{nandini-2019}%
  \BibitemOpen
  \bibfield  {author} {\bibinfo {author} {\bibfnamefont {D.~C.}\ \bibnamefont
  {Ronquillo}}, \bibinfo {author} {\bibfnamefont {A.}~\bibnamefont {Vengal}}, \
  and\ \bibinfo {author} {\bibfnamefont {N.}~\bibnamefont {Trivedi}},\
  }\bibfield  {title} {\enquote {\bibinfo {title} {Signatures of
  magnetic-field-driven quantum phase transitions in the entanglement entropy
  and spin dynamics of the kitaev honeycomb model},}\ }\href {\doibase
  10.1103/PhysRevB.99.140413} {\bibfield  {journal} {\bibinfo  {journal} {Phys.
  Rev. B}\ }\textbf {\bibinfo {volume} {99}},\ \bibinfo {pages} {140413}
  (\bibinfo {year} {2019})}\BibitemShut {NoStop}%
\bibitem [{\citenamefont {Wang}\ \emph {et~al.}(2021)\citenamefont {Wang},
  \citenamefont {Qi}, \citenamefont {Xi}, \citenamefont {Wang}, \citenamefont
  {Yu},\ and\ \citenamefont {Li}}]{wang-2021}%
  \BibitemOpen
  \bibfield  {author} {\bibinfo {author} {\bibfnamefont {S.}~\bibnamefont
  {Wang}}, \bibinfo {author} {\bibfnamefont {Z.}~\bibnamefont {Qi}}, \bibinfo
  {author} {\bibfnamefont {B.}~\bibnamefont {Xi}}, \bibinfo {author}
  {\bibfnamefont {W.}~\bibnamefont {Wang}}, \bibinfo {author} {\bibfnamefont
  {S.-L.}\ \bibnamefont {Yu}}, \ and\ \bibinfo {author} {\bibfnamefont {J.-X.}\
  \bibnamefont {Li}},\ }\bibfield  {title} {\enquote {\bibinfo {title}
  {Comprehensive study of the global phase diagram of the
  $j\ensuremath{-}k\ensuremath{-}\mathrm{\ensuremath{\Gamma}}$ model on a
  triangular lattice},}\ }\href {\doibase 10.1103/PhysRevB.103.054410}
  {\bibfield  {journal} {\bibinfo  {journal} {Phys. Rev. B}\ }\textbf {\bibinfo
  {volume} {103}},\ \bibinfo {pages} {054410} (\bibinfo {year}
  {2021})}\BibitemShut {NoStop}%
\bibitem [{\citenamefont {Nasu}\ \emph {et~al.}(2015)\citenamefont {Nasu},
  \citenamefont {Udagawa},\ and\ \citenamefont {Motome}}]{nasu-motome-2015}%
  \BibitemOpen
  \bibfield  {author} {\bibinfo {author} {\bibfnamefont {J.}~\bibnamefont
  {Nasu}}, \bibinfo {author} {\bibfnamefont {M.}~\bibnamefont {Udagawa}}, \
  and\ \bibinfo {author} {\bibfnamefont {Y.}~\bibnamefont {Motome}},\
  }\bibfield  {title} {\enquote {\bibinfo {title} {Thermal fractionalization of
  quantum spins in a kitaev model: Temperature-linear specific heat and
  coherent transport of majorana fermions},}\ }\href {\doibase
  10.1103/PhysRevB.92.115122} {\bibfield  {journal} {\bibinfo  {journal} {Phys.
  Rev. B}\ }\textbf {\bibinfo {volume} {92}},\ \bibinfo {pages} {115122}
  (\bibinfo {year} {2015})}\BibitemShut {NoStop}%
\bibitem [{\citenamefont {Koga}\ and\ \citenamefont
  {Nasu}(2019)}]{koga-nasu-2019}%
  \BibitemOpen
  \bibfield  {author} {\bibinfo {author} {\bibfnamefont {A.}~\bibnamefont
  {Koga}}\ and\ \bibinfo {author} {\bibfnamefont {J.}~\bibnamefont {Nasu}},\
  }\bibfield  {title} {\enquote {\bibinfo {title} {Residual entropy and spin
  fractionalizations in the mixed-spin kitaev model},}\ }\href {\doibase
  10.1103/PhysRevB.100.100404} {\bibfield  {journal} {\bibinfo  {journal}
  {Phys. Rev. B}\ }\textbf {\bibinfo {volume} {100}},\ \bibinfo {pages}
  {100404} (\bibinfo {year} {2019})}\BibitemShut {NoStop}%
\bibitem [{\citenamefont {Li}\ \emph {et~al.}(2021{\natexlab{b}})\citenamefont
  {Li}, \citenamefont {Zhang}, \citenamefont {Wang}, \citenamefont {Wu},
  \citenamefont {Gao}, \citenamefont {Qu}, \citenamefont {Liu}, \citenamefont
  {Gong},\ and\ \citenamefont {Li}}]{li-2021-ncom}%
  \BibitemOpen
  \bibfield  {author} {\bibinfo {author} {\bibfnamefont {H.}~\bibnamefont
  {Li}}, \bibinfo {author} {\bibfnamefont {H.-K.}\ \bibnamefont {Zhang}},
  \bibinfo {author} {\bibfnamefont {J.}~\bibnamefont {Wang}}, \bibinfo {author}
  {\bibfnamefont {H.-Q.}\ \bibnamefont {Wu}}, \bibinfo {author} {\bibfnamefont
  {Y.}~\bibnamefont {Gao}}, \bibinfo {author} {\bibfnamefont {D.-W.}\
  \bibnamefont {Qu}}, \bibinfo {author} {\bibfnamefont {Z.-X.}\ \bibnamefont
  {Liu}}, \bibinfo {author} {\bibfnamefont {S.-S.}\ \bibnamefont {Gong}}, \
  and\ \bibinfo {author} {\bibfnamefont {W.}~\bibnamefont {Li}},\ }\bibfield
  {title} {\enquote {\bibinfo {title} {Identification of magnetic interactions
  and high-field quantum spin liquid in $\alpha$-rucl3},}\ }\href {\doibase
  10.1038/s41467-021-24257-8} {\bibfield  {journal} {\bibinfo  {journal}
  {Nature Communications}\ }\textbf {\bibinfo {volume} {12}},\ \bibinfo {pages}
  {4007} (\bibinfo {year} {2021}{\natexlab{b}})}\BibitemShut {NoStop}%
\bibitem [{\citenamefont {Mandal}\ \emph
  {et~al.}(2012{\natexlab{b}})\citenamefont {Mandal}, \citenamefont {Shankar},\
  and\ \citenamefont {Baskaran}}]{Mandal_2012}%
  \BibitemOpen
  \bibfield  {author} {\bibinfo {author} {\bibfnamefont {S.}~\bibnamefont
  {Mandal}}, \bibinfo {author} {\bibfnamefont {R.}~\bibnamefont {Shankar}}, \
  and\ \bibinfo {author} {\bibfnamefont {G.}~\bibnamefont {Baskaran}},\
  }\bibfield  {title} {\enquote {\bibinfo {title} {Rvb gauge theory and the
  topological degeneracy in the honeycomb kitaev model},}\ }\href {\doibase
  10.1088/1751-8113/45/33/335304} {\bibfield  {journal} {\bibinfo  {journal}
  {Journal of Physics A: Mathematical and Theoretical}\ }\textbf {\bibinfo
  {volume} {45}},\ \bibinfo {pages} {335304} (\bibinfo {year}
  {2012}{\natexlab{b}})}\BibitemShut {NoStop}%
\bibitem [{\citenamefont {Yamada}(2020)}]{yamada-2020}%
  \BibitemOpen
  \bibfield  {author} {\bibinfo {author} {\bibfnamefont {M.~G.}\ \bibnamefont
  {Yamada}},\ }\bibfield  {title} {\enquote {\bibinfo {title} {Anderson--kitaev
  spin liquid},}\ }\href {\doibase 10.1038/s41535-020-00285-3} {\bibfield
  {journal} {\bibinfo  {journal} {npj Quantum Materials}\ }\textbf {\bibinfo
  {volume} {5}},\ \bibinfo {pages} {82} (\bibinfo {year} {2020})}\BibitemShut
  {NoStop}%
\bibitem [{\citenamefont {Hwang}\ \emph {et~al.}(2022)\citenamefont {Hwang},
  \citenamefont {Go}, \citenamefont {Seong}, \citenamefont {Shibauchi},\ and\
  \citenamefont {Moon}}]{khwang-2022}%
  \BibitemOpen
  \bibfield  {author} {\bibinfo {author} {\bibfnamefont {K.}~\bibnamefont
  {Hwang}}, \bibinfo {author} {\bibfnamefont {A.}~\bibnamefont {Go}}, \bibinfo
  {author} {\bibfnamefont {J.~H.}\ \bibnamefont {Seong}}, \bibinfo {author}
  {\bibfnamefont {T.}~\bibnamefont {Shibauchi}}, \ and\ \bibinfo {author}
  {\bibfnamefont {E.-G.}\ \bibnamefont {Moon}},\ }\bibfield  {title} {\enquote
  {\bibinfo {title} {Identification of a kitaev quantum spin liquid by magnetic
  field angle dependence},}\ }\href {\doibase 10.1038/s41467-021-27943-9}
  {\bibfield  {journal} {\bibinfo  {journal} {Nature Communications}\ }\textbf
  {\bibinfo {volume} {13}},\ \bibinfo {pages} {323} (\bibinfo {year}
  {2022})}\BibitemShut {NoStop}%
\bibitem [{\citenamefont {Pervez}\ and\ \citenamefont
  {Mandal}(2023)}]{pervez2023deciphering}%
  \BibitemOpen
  \bibfield  {author} {\bibinfo {author} {\bibfnamefont {S.~M.}\ \bibnamefont
  {Pervez}}\ and\ \bibinfo {author} {\bibfnamefont {S.}~\bibnamefont
  {Mandal}},\ }\href@noop {} {\enquote {\bibinfo {title} {Deciphering competing
  interactions of kitaev-heisenberg-$\mathrm{\ensuremath{\Gamma}}$ system in
  clusters},}\ } (\bibinfo {year} {2023}),\ \Eprint
  {http://arxiv.org/abs/2306.14839} {arXiv:2306.14839 [cond-mat.str-el]}
  \BibitemShut {NoStop}%
\bibitem [{\citenamefont {Deffner}\ and\ \citenamefont
  {Campbell}(2017)}]{deffner-2017}%
  \BibitemOpen
  \bibfield  {author} {\bibinfo {author} {\bibfnamefont {S.}~\bibnamefont
  {Deffner}}\ and\ \bibinfo {author} {\bibfnamefont {S.}~\bibnamefont
  {Campbell}},\ }\bibfield  {title} {\enquote {\bibinfo {title} {Quantum speed
  limits: from heisenberg’s uncertainty principle to optimal quantum
  control},}\ }\href {\doibase 10.1088/1751-8121/aa86c6} {\bibfield  {journal}
  {\bibinfo  {journal} {Journal of Physics A: Mathematical and Theoretical}\
  }\textbf {\bibinfo {volume} {50}},\ \bibinfo {pages} {453001} (\bibinfo
  {year} {2017})}\BibitemShut {NoStop}%
\bibitem [{\citenamefont {Ness}\ \emph {et~al.}(2022)\citenamefont {Ness},
  \citenamefont {Alberti},\ and\ \citenamefont {Sagi}}]{gal-ness-2022}%
  \BibitemOpen
  \bibfield  {author} {\bibinfo {author} {\bibfnamefont {G.}~\bibnamefont
  {Ness}}, \bibinfo {author} {\bibfnamefont {A.}~\bibnamefont {Alberti}}, \
  and\ \bibinfo {author} {\bibfnamefont {Y.}~\bibnamefont {Sagi}},\ }\bibfield
  {title} {\enquote {\bibinfo {title} {Quantum speed limit for states with a
  bounded energy spectrum},}\ }\href {\doibase 10.1103/PhysRevLett.129.140403}
  {\bibfield  {journal} {\bibinfo  {journal} {Phys. Rev. Lett.}\ }\textbf
  {\bibinfo {volume} {129}},\ \bibinfo {pages} {140403} (\bibinfo {year}
  {2022})}\BibitemShut {NoStop}%
\bibitem [{\citenamefont {Mohan}\ and\ \citenamefont {Pati}(2022)}]{pati-2022}%
  \BibitemOpen
  \bibfield  {author} {\bibinfo {author} {\bibfnamefont {B.}~\bibnamefont
  {Mohan}}\ and\ \bibinfo {author} {\bibfnamefont {A.~K.}\ \bibnamefont
  {Pati}},\ }\bibfield  {title} {\enquote {\bibinfo {title} {Quantum speed
  limits for observables},}\ }\href {\doibase 10.1103/PhysRevA.106.042436}
  {\bibfield  {journal} {\bibinfo  {journal} {Phys. Rev. A}\ }\textbf {\bibinfo
  {volume} {106}},\ \bibinfo {pages} {042436} (\bibinfo {year}
  {2022})}\BibitemShut {NoStop}%
\bibitem [{\citenamefont {Aggarwal}\ \emph {et~al.}(2022)\citenamefont
  {Aggarwal}, \citenamefont {Banerjee}, \citenamefont {Ghosh},\ and\
  \citenamefont {Mukhopadhyay}}]{arindam-2022}%
  \BibitemOpen
  \bibfield  {author} {\bibinfo {author} {\bibfnamefont {S.}~\bibnamefont
  {Aggarwal}}, \bibinfo {author} {\bibfnamefont {S.}~\bibnamefont {Banerjee}},
  \bibinfo {author} {\bibfnamefont {A.}~\bibnamefont {Ghosh}}, \ and\ \bibinfo
  {author} {\bibfnamefont {B.}~\bibnamefont {Mukhopadhyay}},\ }\bibfield
  {title} {\enquote {\bibinfo {title} {Non-uniform magnetic field as a booster
  for quantum speed limit: faster quantum information processing},}\ }\href
  {\doibase 10.1088/1367-2630/ac84f9} {\bibfield  {journal} {\bibinfo
  {journal} {New Journal of Physics}\ }\textbf {\bibinfo {volume} {24}},\
  \bibinfo {pages} {085001} (\bibinfo {year} {2022})}\BibitemShut {NoStop}%
\bibitem [{\citenamefont {Wang}\ \emph {et~al.}(2017)\citenamefont {Wang},
  \citenamefont {Dong}, \citenamefont {Yu},\ and\ \citenamefont
  {Li}}]{alternative_6_site}%
  \BibitemOpen
  \bibfield  {author} {\bibinfo {author} {\bibfnamefont {W.}~\bibnamefont
  {Wang}}, \bibinfo {author} {\bibfnamefont {Z.-Y.}\ \bibnamefont {Dong}},
  \bibinfo {author} {\bibfnamefont {S.-L.}\ \bibnamefont {Yu}}, \ and\ \bibinfo
  {author} {\bibfnamefont {J.-X.}\ \bibnamefont {Li}},\ }\bibfield  {title}
  {\enquote {\bibinfo {title} {Theoretical investigation of magnetic dynamics
  in $\ensuremath{\alpha}\ensuremath{-}{\mathrm{rucl}}_{3}$},}\ }\href
  {\doibase 10.1103/PhysRevB.96.115103} {\bibfield  {journal} {\bibinfo
  {journal} {Phys. Rev. B}\ }\textbf {\bibinfo {volume} {96}},\ \bibinfo
  {pages} {115103} (\bibinfo {year} {2017})}\BibitemShut {NoStop}%
\bibitem [{\citenamefont {Maity}\ \emph {et~al.}(2020)\citenamefont {Maity},
  \citenamefont {Iqbal},\ and\ \citenamefont {Mandal}}]{atanufisher}%
  \BibitemOpen
  \bibfield  {author} {\bibinfo {author} {\bibfnamefont {A.}~\bibnamefont
  {Maity}}, \bibinfo {author} {\bibfnamefont {Y.}~\bibnamefont {Iqbal}}, \ and\
  \bibinfo {author} {\bibfnamefont {S.}~\bibnamefont {Mandal}},\ }\bibfield
  {title} {\enquote {\bibinfo {title} {Competing orders in a frustrated
  heisenberg model on the fisher lattice},}\ }\href {\doibase
  10.1103/PhysRevB.102.224404} {\bibfield  {journal} {\bibinfo  {journal}
  {Phys. Rev. B}\ }\textbf {\bibinfo {volume} {102}},\ \bibinfo {pages}
  {224404} (\bibinfo {year} {2020})}\BibitemShut {NoStop}%
\bibitem [{\citenamefont {d'Ornellas}\ and\ \citenamefont
  {Knolle}(2024)}]{knollestar}%
  \BibitemOpen
  \bibfield  {author} {\bibinfo {author} {\bibfnamefont {P.}~\bibnamefont
  {d'Ornellas}}\ and\ \bibinfo {author} {\bibfnamefont {J.}~\bibnamefont
  {Knolle}},\ }\bibfield  {title} {\enquote {\bibinfo {title}
  {Kitaev-heisenberg model on the star lattice: From chiral majorana fermions
  to chiral triplons},}\ }\href {\doibase 10.1103/PhysRevB.109.094421}
  {\bibfield  {journal} {\bibinfo  {journal} {Phys. Rev. B}\ }\textbf {\bibinfo
  {volume} {109}},\ \bibinfo {pages} {094421} (\bibinfo {year}
  {2024})}\BibitemShut {NoStop}%
\end{thebibliography}%
\section{Appendix}
\label{appendix}
\appendix
\newcounter{defcounter}
\setcounter{defcounter}{0}
\section{Finite $\Gamma$ results}{\label{App:A_2}}
\begin{figure*}[h]
	\includegraphics[width=1.9\columnwidth,height=!]{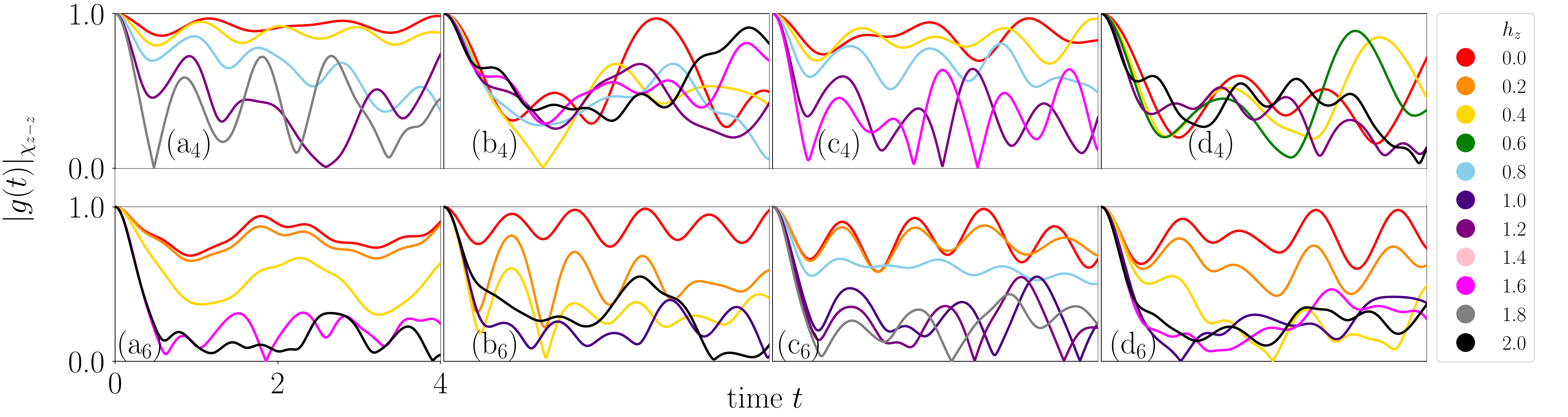}
	\caption{Gauge Majorana dynamics for $J=0$ and (a$_N$) $K=1.0,~\Gamma=0.5$, (b$_N$) $K=-1.0,~\Gamma=0.5$, (c$_N$) $K=1.0,~\Gamma=-0.5$, (d$_N$) $K=-1.0,~\Gamma=-0.5$. Different strength of $h_z$ are given in legend box.}
	\label{gauge_finite_Gamma}
\end{figure*}
\begin{figure*}[h]
	\includegraphics[width=1.9\columnwidth,height=!]{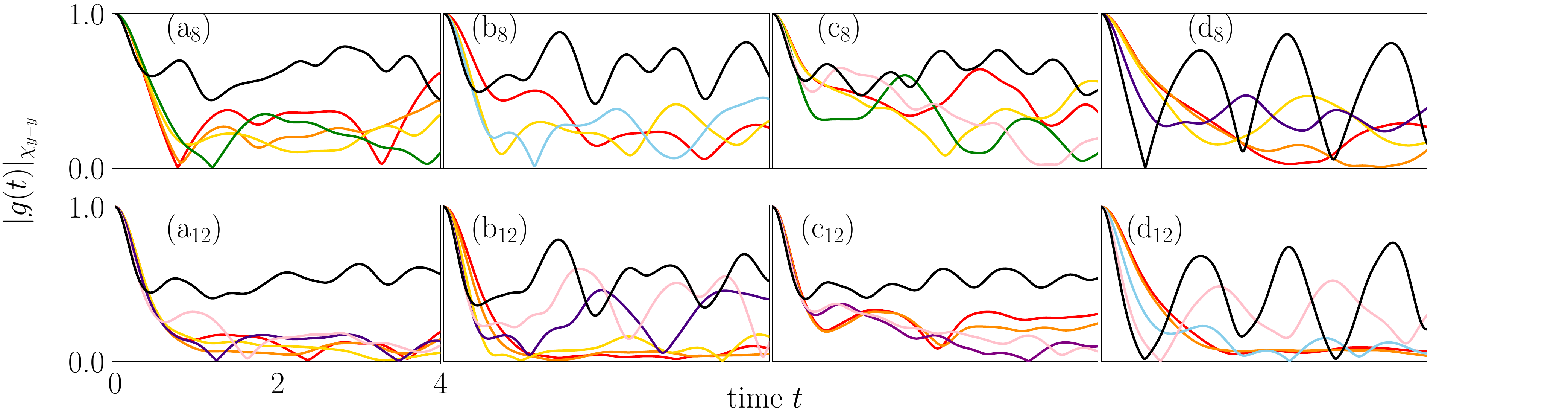}
	\caption{Gauge Majorana dynamics for $J=0$ and (a$_N$) $K=1.0,~\Gamma=0.5$, (b$_N$) $K=-1.0,~\Gamma=0.5$, (c$_N$) $K=1.0,~\Gamma=-0.5$, (d$_N$) $K=-1.0,~\Gamma=-0.5$. Different strength of $h_z$ are given in legend box of FIG.\ref{gauge_finite_Gamma}.}
	\label{gauge_finite_Gamma1}
\end{figure*}
\begin{figure*}[h]
	\includegraphics[width=1.9\columnwidth,height=!]{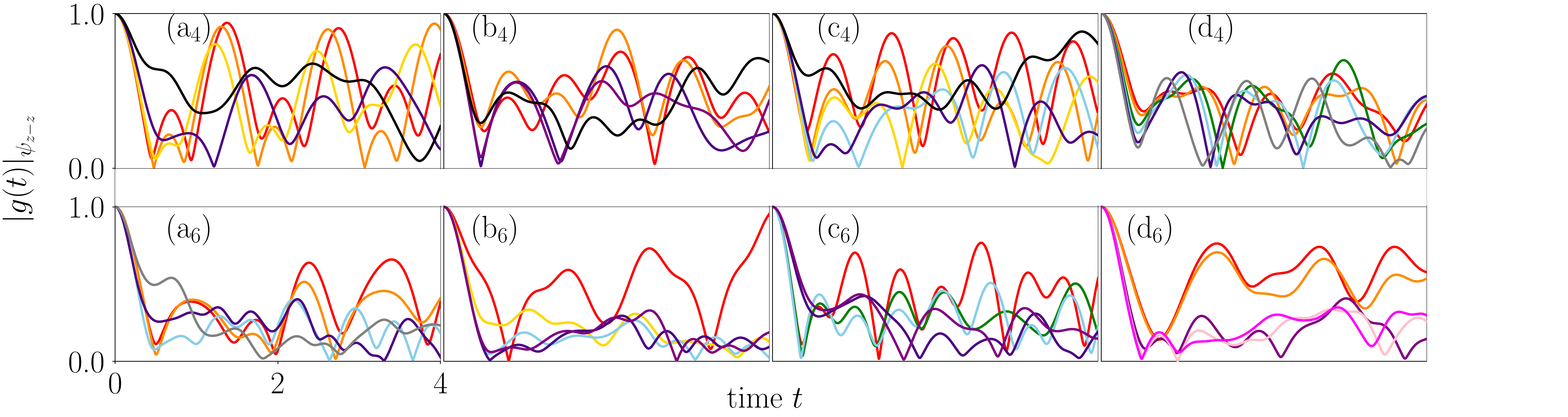}
	\caption{Matter Majorana dynamics for $J=0$ and (a$_N$) $K=1.0,~\Gamma=0.5$, (b$_N$) $K=-1.0,~\Gamma=0.5$, (c$_N$) $K=1.0,~\Gamma=-0.5$, (d$_N$) $K=-1.0,~\Gamma=-0.5$. Different strength of $h_z$ are given in legend box of FIG.\ref{gauge_finite_Gamma}.}
	\label{matter_finite_Gamma}
\end{figure*}
\begin{figure*}[h]
	\includegraphics[width=1.9\columnwidth,height=!]{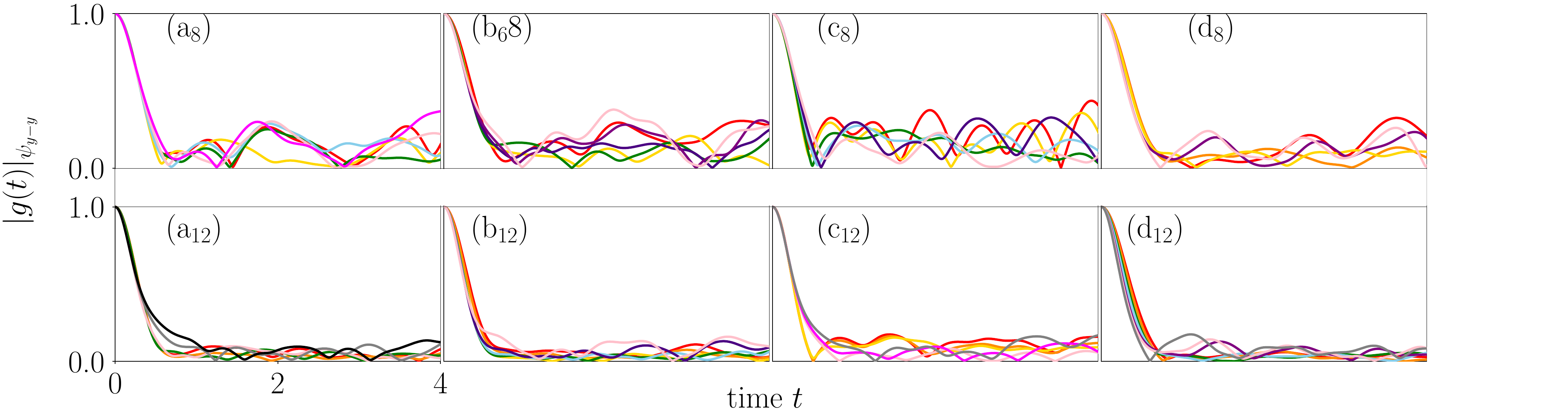}
	\caption{Matter Majorana dynamics for $J=0$ and (a$_N$) $K=1.0,~\Gamma=0.5$, (b$_N$) $K=-1.0,~\Gamma=0.5$, (c$_N$) $K=1.0,~\Gamma=-0.5$, (d$_N$) $K=-1.0,~\Gamma=-0.5$. Different strength of $h_z$ are given in legend box of FIG.\ref{gauge_finite_Gamma}.}
	\label{matter_finite_Gamma1}
\end{figure*}
\end{document}